\documentclass[aps, prc,
reprint,
longbibliography, 
nofootinbib,
superscriptaddress]{revtex4-1}

\makeatletter
\newcommand*\mysize{%
  \normalsize
}
\makeatother

\makeatletter
\def\@bibdataout@aps{%
\immediate\write\@bibdataout{%
@CONTROL{%
apsrev41Control%
\longbibliography@sw{%
    ,author="08",editor="1",pages="1",title="0",year="1"%
    }{%
    ,author="08",editor="1",pages="1",title="",year="1"%
    }%
  }%
}%
\if@filesw \immediate \write \@auxout {\string \citation {apsrev41Control}}\fi
}
\makeatother

\usepackage{setspace}

\usepackage{amsfonts}
\usepackage{amsmath}
\usepackage{amssymb}

\DeclareFontFamily{OT1}{pzc}{}
\DeclareFontShape{OT1}{pzc}{m}{it}{<-> s * [1.10] pzcmi7t}{}
\DeclareMathAlphabet{\mathpzc}{OT1}{pzc}{m}{it}

\usepackage{booktabs}
\AtBeginDocument{
\heavyrulewidth=.08em
\lightrulewidth=.05em
\cmidrulewidth=.03em
\belowrulesep=.65ex
\belowbottomsep=0pt
\aboverulesep=.4ex
\abovetopsep=0pt
\cmidrulesep=\doublerulesep
\cmidrulekern=.5em
\defaultaddspace=.5em
}

\usepackage{cancel}
\usepackage{caption}
\usepackage{chemformula} 
\usepackage{comment} 
\usepackage[inline]{enumitem}
\usepackage[nowatermark]{fixmetodonotes}
\usepackage{graphicx}
\usepackage{isotope}
\usepackage{makecell} 
\usepackage{mathtools} 
\usepackage{multirow} 
\usepackage{siunitx}
\usepackage{stackrel} 
\DeclareSIUnit\ton{t}
\DeclareSIUnit\parsec{pc}
\DeclareSIUnit[number-unit-product = ]\percent{\char`\%}

\usepackage{threeparttable} 
\usepackage{xifthen} 

\usepackage[colorlinks, urlcolor=blue, citecolor=., menucolor=.,
  anchorcolor=., linkcolor=., runcolor=.]{hyperref}

\usepackage{cleveref}
\crefrangeformat{equation}{eqs. #3(#1)#4--#5(#2)#6}

\crefformat{footnote}{#2\textsuperscript{#1}#3}

\DeclareMathOperator{\Tr}{Tr}

\newcommand{\fragmentKinELab}{ \varepsilon_\text{lab} }
\newcommand{\FermiEDiff}{ \mathcal{E} }
\newcommand{\NucleonSpecies}{ \mathpzc{N} }
\newcommand{\fragmentMomCM}{ \mathpzc{k} }
\newcommand{\SMatrixElement}{\left< S_{\ell j} \right>}

\newcommand{\fragment}{ \ensuremath{a} }

\newcommand{\FluxAvgTotXSec}{ \left<\sigma\right> }

\newcommand{\Eavg}{ \left<E_\nu\right> }

\begin{document}


\newcommand{\marley}{\texttt{MARLEY}}

\newcommand{\version}{1.2.0}

\newcommand{\unaryminus}{\scalebox{0.5}[0.72]{\( - \!\)}}

\newcommand{\pnu}{k}
\newcommand{\plep}{\pnu^\prime}
\newcommand{\pNi}{p}
\newcommand{\pNf}{\pNi^\prime}

\newcommand{\Lep}{\mathsf{L}}

\newcommand{\Nuc}{\mathsf{W}}

\newcommand{\NucMat}{\mathcal{N}}

\newcommand{\nucIndex}{n}

\newcommand{\FermiOp}{\mathcal{O}_\mathrm{F}}

\newcommand{\FermiMat}{B(\mathrm{F})}

\newcommand{\GTOp}{\mathcal{O}_\mathrm{GT}}

\newcommand{\GTMat}{B(\mathrm{GT})}

\newcommand{\CoulombFactor}{F_C}

\newcommand{\FNRpLep}{\mathpzc{K}}
\newcommand{\FNReLep}{\mathpzc{E}}
\newcommand{\FNRpLepEff}{\FNRpLep_\text{\,\,eff}}
\newcommand{\FNReLepEff}{\FNReLep_\text{\,eff}}

\newcommand{\BFf}{\ensuremath{B_f(F)}}
\newcommand{\BGTf}{\ensuremath{B_f(GT)}}

\newcommand{\Sel}{ S^J_{\ell j \alpha,\, \ell j \alpha} }
\newcommand{\Sinel}{ S^J_{\ell^\prime j^\prime \beta,\, \ell j \alpha} }
\newcommand{\SelSpinZero}{\left< S^j_{\ell j \alpha,\, \ell j \alpha} \right>}
\newcommand{\totKinE}{\varepsilon}
\newcommand{\totKinEmax}{\varepsilon_\text{max}}
\newcommand{\totWidth}{\Gamma}

\title{Nuclear de-excitations in low-energy
charged-current $\nu_e$ scattering on \isotope[40]{Ar}}

\author{Steven Gardiner}
\email{gardiner@fnal.gov}
\affiliation{Fermi National Accelerator Laboratory, Batavia,
Illinois 60510 USA}
\affiliation{Department of Physics, University of California, Davis,
California 95616 USA}

\date{\today}

\begin{abstract}
\begin{description}
\item[Background] Large argon-based neutrino detectors, such as those planned
for the Deep Underground Neutrino Experiment, have the potential to
provide unique sensitivity to low-energy (few to tens of MeV) electron
neutrinos produced by core-collapse supernovae. Despite their importance for
neutrino energy reconstruction, nuclear de-excitations following
charged-current $\nu_e$ absorption on \isotope[40]{Ar} have never been studied
in detail at supernova energies.
\item[Purpose] I develop a model of nuclear de-excitations that occur following
the $\isotope[40]{Ar}(\nu_e,e^{-})\isotope[40]{K}^*$ reaction. This model is
applied to the calculation of exclusive cross sections.
\item[Methods] A simple expression for the inclusive differential cross section
is derived under the allowed approximation. Nuclear de-excitations are
described using a combination of measured $\gamma$-ray decay schemes and the
Hauser-Feshbach statistical model. All calculations are carried out using a
novel Monte Carlo event generator called \texttt{MARLEY} (Model of Argon
Reaction Low Energy Yields).
\item[Results] Various total and differential cross sections are presented. Two
de-excitation modes, one involving only $\gamma$-rays and the other including
single neutron emission, are found to be dominant at few tens-of-\si{\MeV}
energies.
\item[Conclusions] Nuclear de-excitations have a strong impact on the
achievable energy resolution for supernova $\nu_e$ detection in liquid argon.
Tagging events involving neutron emission, though difficult, could
substantially improve energy reconstruction. Given a suitable calculation of
the inclusive cross section, the \marley\ nuclear de-excitation model may
readily be applied to other scattering processes.
\end{description}
\end{abstract}


\maketitle

\section{Introduction}

Core-collapse supernovae are exceptionally intense sources of tens-of-\si{\MeV}
neutrinos and antineutrinos of all flavors. In a typical supernova, about
\num{e58} neutrinos are released in a burst lasting tens of seconds. Although
the first observation of supernova neutrinos by the Kamiokande-II
\cite{Hirata1987}, Baksan \cite{Alekseev1987}, and Irvine-Michigan-Brookhaven \cite{Bionta1987}
detectors in 1987 yielded a total of only two dozen events, the scientific
impact of this dataset has been tremendous, leading to numerous publications on
a wide variety of subjects
\cite{Raffelt1990,Schaeffer1990,Jegerlehner1996,Vissani2014,Mirizzi2016,Branch2017}.
In the years since first detection, a worldwide network of large neutrino
experiments, most built primarily for other applications, stands ready to
perform a second, high-statistics measurement if a core-collapse supernova
should occur within the galaxy \cite{Scholberg2012}. Due to the slow rate of
galactic core-collapse supernovae (estimated to be about 1.6 per century
\cite{Rozwadowska2021}), the prospect of such a measurement represents a rare
but valuable opportunity to shed light on the details of core-collapse and
nucleosynthesis, study neutrinos under extreme conditions, search for evidence
of physics beyond the Standard Model, and explore many other topics
\cite{Muller2019,Horiuchi2018,Raffelt2011}.

A full realization of the physics potential of the next galactic core-collapse
supernova will require a simultaneous measurement of neutrinos of all flavors.
While detectors based on water and hydrocarbon scintillator will primarily
detect $\bar{\nu}_e$ via inverse beta decay (IBD) \begin{equation} \bar{\nu}_e
+ p \rightarrow e^{+} + n \,, \end{equation} liquid-argon-based detectors are
anticipated to provide unique sensitivity \cite{Abi2020,Ankowski2016} to
$\nu_e$ via the charged-current (CC) reaction \begin{equation}
\label{eq:nueArCC} \nu_e + \isotope[40]{Ar} \rightarrow e^{-} +
\isotope[40]{K}^{*} \end{equation} which dominates the expected signal at
supernova energies.

Within the decade, the Deep Underground Neutrino Experiment (DUNE) will begin
operating four ten-kiloton liquid argon time projection chambers
(\mbox{LArTPCs}) with the primary goals of studying long-baseline oscillations
of accelerator neutrinos, searching for proton decay, and measuring the $\nu_e$
flux from a galactic core-collapse supernova if one should occur during the
lifetime of the experiment \cite{DUNETDRvol1}. Initial studies of the
sensitivity of DUNE to supernova neutrinos, performed by the collaboration
itself \cite{Abi2020} and by smaller groups (e.g., ref.~\cite{Li2021}) show
promise, and the potential exists for measurements by DUNE of other low-energy
neutrinos, notably those produced by the Sun \cite{DUNEsolar}.

In addition to DUNE, three sub-kiloton LArTPCs, SBND
\cite{Tufanli2017,Brailsford2017}, MicroBooNE \cite{Acciarri2017}, and ICARUS
\cite{Amerio2004}, are currently operating or being installed in the Booster
Neutrino Beam at Fermilab. A joint effort between the three experimental
collaborations, known as the Short Baseline Neutrino (SBN) program
\cite{Machado2019,Acciarri2015}, will perform precision measurements of
neutrino oscillations. In addition to this primary mission, the SBN detectors
will pursue a variety of other physics measurements and are expected to be
sensitive to supernova neutrinos. To ensure that data from a core-collapse
supernova would be fully recorded over the ten-second burst, the
MicroBooNE collaboration operates a first-of-its-kind continuous readout stream
and has demonstrated its capabilities via reconstruction of Michel electrons
produced by decays of cosmic muons \cite{Abratenko2020}.

While much remains to be done to fully exploit the low-energy capabilities of
LArTPCs, a first demonstration by the ArgoNeuT \cite{Anderson2012} experiment
of reconstruction of \si{\MeV}-scale activity due to
accelerator-neutrino-induced neutrons and de-excitation $\gamma$-rays achieved
a detection threshold of around \SIrange{200}{300}{\keV} \cite{ArgoneutMeV}.
These encouraging initial results have prompted further experimental work by
MicroBooNE \cite{PublicNoteMeV} and multiple simulation-based studies
considering the implications for reconstruction of both high- and low-energy
physics events \cite{Friedland2019,Castiglioni2020}.

In future analyses of supernova neutrino data, the event-by-event reconstructed
neutrino energy obtained by each detector will be of primary interest. For IBD
events in water or scintillator, because only a single hadronic final state (a
free neutron) is accessible at tens-of-\si{\MeV} energies, a measurement of the
outgoing positron energy is sufficient to reconstruct the antineutrino energy
with high accuracy. Due to the use of a complex nuclear target (argon) in
LArTPCs, however, various nuclear transitions may occur in response to CC
$\nu_e$ absorption, and thus a one-to-one mapping (up to nuclear recoil)
between the neutrino and electron energies no longer exists.

To fully reconstruct the neutrino energy in the argon case, the reaction
Q-value, i.e., the energy imparted to the nuclear transition, must be inferred
by detecting the nuclear de-excitation products. For transitions to bound
nuclear energy levels, the neutrino energy is in principle fully recoverable by
measuring the energies of all de-excitation $\gamma$-rays in addition to the
primary electron. For transitions to unbound nuclear states, a model is needed
to correct for missing energy associated with undetected nuclear fragments. In
practice, an experimental analysis that attempts to isolate the simpler bound
transitions will also need a detailed de-excitation model in order to estimate
the purity of the event selection.

With the exception of a recent first measurement \cite{Akimov2021} of coherent
elastic neutrino-nucleus scattering, experimental data have not yet
been obtained for neutrino-argon cross sections in the supernova energy regime.
Furthermore, only a few measurements with limited precision are available for
low-energy inelastic neutrino scattering on any nuclear target \cite[table
3]{Ajimura2017}. Despite this, a substantial literature exists for theoretical
calculations of the $\isotope[40]{Ar}(\nu_e,e^{-})\isotope[40]{K}^{*}$ process.
A review through 2018 is provided in ref.~\cite[sec. 7.1]{Gardiner2018}, with a
notable recent addition being two publications \cite{CRPA1,CRPA2} which employ
a Continuum Random Phase Approximation (CRPA) model to study this cross section
above the nucleon emission threshold.

While highly useful for providing competing estimates of event rates in DUNE
and other argon-based detectors, all published calculations for this cross
section to date share the limitation of being fully inclusive, i.e.,
predictions are made that consider only the kinematics of the outgoing
electron. At very low neutrino energies, where only transitions to bound
nuclear states are possible, this is not problematic: measured de-excitation
$\gamma$-ray branching ratios exist for many levels of the daughter
\isotope[40]{K} nucleus, and missing data may be addressed using
straightforward theoretical estimates. However, above
about \SIrange{15}{20}{\MeV}, kinematic access to unbound nuclear states
becomes appreciable, and a detailed treatment of the competition between
various de-excitation channels (including emission of both $\gamma$-rays and
nuclear fragments) is needed.

Although such a treatment has not previously been provided for
\isotope[40]{Ar}, detailed modeling of nuclear de-excitations induced by
low-energy neutrino interactions has been pursued for a number of other nuclei
\cite{Kolbe1992,Langanke1996,Kolbe2001,Cheoun2012b,Bandyopadhyay2017,Vale2016}.
A universal assumption made by all of these calculations (as well as the
present work) is that of \textit{compound nucleus} formation: following the
primary interaction, the nucleus is left in a thermally-equilibrated excited
state that decays independently of the details of its formation process. While
further work is needed to fully investigate the adequacy of this assumption for
low-energy neutrino-nucleus reactions, both theoretical
calculations~\cite{Kim2009,Cheoun2011a} and electron scattering
data~\cite{Flowers1978} suggest that compound processes dominate over the
direct nucleon knock-out important at higher energies.

In this paper, I present the first calculations at tens-of-\si{\MeV} energies
for cross sections for exclusive final states of the reaction
$\isotope[40]{Ar}(\nu_e,e^{-})\isotope[40]{K}^{*}$, emphasizing the role of
nuclear de-excitation processes. In \cref{sec:xsec_model}, I develop a simple
model for the inclusive differential cross section, relying on approximations
that work best at low momentum transfers. The derivation in
\cref{sec:xsec_model} fully determines the cross section up to the values of
two (Fermi and Gamow-Teller) nuclear matrix elements, $\FermiMat$ and $\GTMat$, which are considered in
\cref{sec:allowed_MEs}. While relevant neutrino scattering data are currently
unavailable, the needed values of these matrix elements at low excitation
energies may be extracted from measurements of related processes. I supplement
these measurements with the results of a theoretical calculation at high
excitation energies to obtain a full treatment of the inclusive cross section.
In \cref{sec:deex_model}, I describe a detailed model of nuclear de-excitations
that can be used together with the inclusive cross section to obtain
predictions for exclusive final states. In \cref{sec:results}, I present
example results calculated using the models developed in the previous sections.

To enable practical calculations that have already helped to inform studies of
DUNE's sensitivity to supernova neutrinos \cite{Abi2020}, all of the physics
models described herein have been implemented in a new Monte Carlo event
generator called \marley\ (Model of Argon Reaction Low Energy Yields). All
results shown in this work may be reproduced using version \version\ of
\marley\ \cite{MARLEYv1.2.0}, which is publicly available as an open-source
software project \cite{MARLEYWebsite}. Documentation of the technical details
of \marley\ and usage instructions are available in ref.~\cite{marleyCPC}.

Due to the compound nucleus assumption, the \marley\ de-excitation model may
easily be used in the future to provide exclusive predictions for a more
refined calculation of the inclusive CC $\nu_e$ absorption cross section for
\isotope[40]{Ar}. A similar approach to modeling de-excitations for other
reaction modes and target nuclei is likewise possible, and I welcome potential
collaboration on this topic. Prospects for improving \marley\ predictions
beyond the proof-of-concept reported here are briefly considered in
\cref{sec:conclusions}.

\section{Inclusive cross section model}
\label{sec:xsec_model}

For momentum transfers that are small compared to the $W$ boson mass, the
tree-level amplitude $\mathcal{M}$ for inclusive charged-current
neutrino-nucleus scattering may be represented diagramatically as
\begin{equation}
i\mathcal{M} = \vcenter{\hbox{ \includegraphics{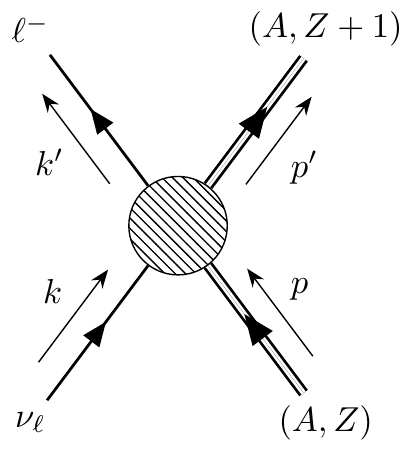} }}.
\end{equation}
The corresponding differential cross section may be written in the form
\begin{equation}
\label{eq:xsec1}
\frac{ d\sigma }{ dQ^2 }
= \frac{ G_F^2 \, |V_{\mathrm{ud}}|^2 }{ 32\, \pi \, (s - m_i^2)^2 }
\, \CoulombFactor \, \Lep_{\mu\nu} \, \Nuc^{\mu\nu} \,,
\end{equation}
where $q = \pnu - \plep = \pNf - \pNi$ is the four-momentum transfer, $Q^2
\equiv -q^2$, $G_F$ is the Fermi constant, $V_\mathrm{ud}$ is the
Cabibbo–Kobayashi–Maskawa matrix element connecting the up and down quarks,
Mandelstam $s$ is the square of the total energy in the center-of-momentum (CM)
frame, and $m_i$ is the mass of the initial-state nucleus.
Discussion of the Coulomb correction factor $\CoulombFactor$
is deferred to \cref{sec:Coulomb}.

The leptonic ($\Lep_{\mu\nu}$) and hadronic ($\Nuc^{\mu\nu}$) tensors are
defined by
\begin{align}
\label{eq:CC_lepton_tensor}
\Lep_{\mu\nu} &\equiv \Tr[\gamma_\mu \, (1 - \gamma_5) \, \cancel{\pnu} \,
\gamma_{\nu} \, (1 - \gamma_5) \, (\cancel{\pnu}^\prime + m_\ell)]
\\ &= 8\left[ \pnu_\mu \, \plep_\nu
+ \pnu_\nu \, \plep_\mu
- g_{\mu \nu}\, (\pnu \cdot \plep)
- i\epsilon_{\mu \nu \rho \sigma} \, {\pnu}^\rho {\plep}^\sigma
\right]
\end{align}
and
\begin{equation}
\label{eq:CC_nuclear_tensor}
\Nuc^{\mu\nu} \equiv \frac{1}{2J_i + 1} \sum_{M_i} \sum_{M_f}
\NucMat^\mu\,\NucMat^{\nu*}.
\end{equation}
Here $m_\ell$ is the mass of the final-state lepton, $J_i$ ($J_f$) is the
initial (final) nuclear spin, and $M_i$ ($M_f$) is the third component of the
nuclear spin in the initial (final) state.

Under the impulse approximation, the nuclear matrix element may be written in
coordinate space as
\begin{equation}
\label{eq:nuclear_matrix_element}
\NucMat^\mu = \big< f \big| \, {\textstyle\sum_{\nucIndex=1}^A} \, e^{ i \mathbf{q}
\cdot \mathbf{x}(\nucIndex) } \, j^\mu(\nucIndex) \, \big| i \big>
\end{equation}
where $\mathbf{q}$ is the three-momentum transfer and the sum runs over all $A$
nucleons. The weak current operator $j^\mu(\nucIndex)$ is understood to act only
on the $\nucIndex$th nucleon, as is the position operator
$\mathbf{x}(\nucIndex)$. The state vectors in \cref{eq:nuclear_matrix_element}
are normalized relativistically, i.e.,
\begin{align}
\big< i \big| i \big> &= 2 E_i
&
\big< f \big| f \big> &= 2 E_f
\end{align}
where $E_i$ ($E_f$) is the total energy of the nucleus in the initial (final)
state. \Cref{eq:xsec1} contains an implied sum over the accessible final
nuclear states.

\subsection{Allowed approximation}
\label{sec:allowed_approx}

The full expression for the single-nucleon weak current operator $j^\mu$ is
well-known and is given in ref.~\cite{Gardiner2018} among other places. For this
study, however, I evaluate the current operator in the \textit{allowed
approximation}, which combines the long-wavelength ($q \to 0$) limit and the
slow-nucleon limit (in which the momentum of the initial struck nucleon is
neglected compared to its mass).

Under this approximation, the weak current operator reduces to the simple form
\begin{align}
j^0 &= g_V \, t_- & j^a &= -\sigma^a \, g_A \, t_-
\end{align}
where $j^0$ is the time component and the three Cartesian spatial components
are labeled with $a \in \{1,2,3\}$. The
time component of the nuclear matrix element $\NucMat^\mu$
is given by
\begin{equation}
\label{eq:NucMat0}
\NucMat^{0} = \frac{ g_V }{ \sqrt{2J_i + 1} } \, \delta_{J_i \, J_f}
\, \delta_{M_i \, M_f} \, \big< f \big\Vert \FermiOp
\big\Vert i \big>
\end{equation}
while the spatial components may be written in spherical
coordinates as
\begin{equation}
\label{eq:NucMatB}
\NucMat^{w} = \frac{ -g_A\, (-1)^{J_i - M_i} }{ \sqrt{3} }
 \, \big( J_f \; M_f \; J_i \; {-M_i} \, \big| \, 1 \; w \big)
\, \big< f \big\Vert \GTOp
\big\Vert i \big>
\end{equation}
where $w \in \{-1,0,1\}$ and $g_V$ ($g_A$) is the vector (axial-vector) weak
coupling constant of the nucleon. The Fermi (F) and Gamow-Teller (GT) operators are
defined by
\begin{align}
\FermiOp &\equiv \sum_{\nucIndex=1}^A t_-(\nucIndex) \\[0.5\baselineskip]
\GTOp &\equiv \sum_{\nucIndex=1}^A \boldsymbol{\sigma}(\nucIndex) \, t_-(\nucIndex)
\end{align}
where $\boldsymbol{\sigma}$ is the Pauli vector, and $t_-$, the isospin-lowering
operator, converts a neutron into a proton. Double bars ($\Vert$) denote matrix
elements which have been reduced via the Wigner-Eckhart theorem.

\Cref{eq:NucMat0,eq:NucMatB} may be used to evaluate the elements of the
hadronic tensor $\Nuc^{\mu\nu}$. Under the allowed approximation, these
become
\begin{align}
\Nuc^{00} &= 4 \, E_i \, E_f \, \FermiMat
\\[0.5\baselineskip]
\Nuc^{ab} &= \frac{4}{3} \, \delta_{ab} \, E_i \, E_f \, \GTMat
\\[0.5\baselineskip]
\Nuc^{0a} &= \Nuc^{a0} = 0
\end{align}
where the reduced Fermi and Gamow-Teller matrix elements are given by
\begin{align}
\label{eq:BF}
\FermiMat &\equiv \frac{ g_V^2 }{ 2J_i + 1 } \Big|\big< J_f \, \big\lVert
\, \FermiOp \, \big\rVert \, J_i \big> \Big|^2
\\[0.5\baselineskip]
\label{eq:BGT}
\GTMat &\equiv \frac{ g_A^2 }{ 2J_i + 1 } \Big|\big< J_f \, \big\lVert
\, \GTOp \, \big\rVert \, J_i \big> \Big|^2 \,.
\end{align}
The state vectors labeled using the nuclear spin ($J_i$ or $J_f$) are
normalized to unity:
\begin{equation}
\big< J_i \big| J_i \big> =
\big< J_f \big| J_f \big> = 1.
\end{equation}
The reduced matrix elements satisfy the spin-parity selection rules
\begin{equation}
\label{eq:Fermi_sel_rule}
B(\mathrm{F}) = 0 \text{ unless } J_f = J_i \text{ and } \Pi_f = \Pi_i
\end{equation}
and
\begin{equation}
\label{eq:GT_sel_rule}
B(\mathrm{GT}) = 0 \text{ unless } |J_i - 1| \leq J_f \leq J_i + 1
\text{ and } \Pi_f = \Pi_i\,.
\end{equation}
where $\Pi_i$ ($\Pi_f$) is the initial (final) nuclear parity.

Combining the results above leads to the following expression for the
allowed approximation differential cross section in the CM frame:
\begin{widetext}
\begin{align}
\label{eq:xsec2}
\frac{ d\sigma }{d\cos \theta_{\ell}} = &
\; \frac{ G_F^2 \, |V_{\mathrm{ud}}|^2 }{ 2 \, \pi } \,
\CoulombFactor \, \bigg[ \frac{ E_i \, E_f }{ s } \bigg]
\, E_\ell \left|\mathbf{p}_\ell\right|
\bigg[ \Big(1 + \beta_\ell \cos \theta_{\ell} \Big)\,
\FermiMat
+\, \Big(1 - \frac{1}{3} \, \beta_\ell \cos \theta_{\ell} \Big)
\, \GTMat \bigg].
\end{align}
\end{widetext}
Here $E_\ell$, $\mathbf{p}_\ell$, $\theta_\ell$, and $\beta_\ell = E_\ell /
|\mathbf{p}_\ell| $ are, respectively, the total energy, three-momentum, scattering
angle, and speed of the final-state lepton. The factor $E_i \, E_f / s$ accounts
for nuclear recoil and is commonly neglected.

In the CM frame, the particle energies are independent of the scattering
angle $\theta_\ell$. As a result, integration of the total cross section
is trivial and leads to the expression
\begin{equation}
\label{eq:tot_xsec}
\sigma = \frac{ G_F^2 \, |V_{\mathrm{ud}}|^2 }{ \pi } \,
\CoulombFactor \,
\bigg[ \frac{ E_i \, E_f }{ s } \bigg]
\, E_\ell \left|\mathbf{p}_\ell\right|
\bigg[ \FermiMat + \GTMat \bigg].
\end{equation}
As was the case for \cref{eq:xsec1}, the cross-section formulas in
\cref{eq:xsec2,eq:tot_xsec} contain an implicit sum over nuclear final states.

\subsection{Coulomb corrections}
\label{sec:Coulomb}

Final-state interactions (FSIs) of the outgoing charged lepton with the Coulomb
field of the nucleus have a significant effect on the cross section at low
energies. While a detailed treatment of Coulomb FSIs is achievable via the
distorted-wave Born approximation, a much more convenient approximation scheme
based on the work of Engel \cite{Engel1998} is typically used, e.g., in
refs.~\cite{Volpe2002,Ydrefors2012a,CRPA1}.

Under this approach, the Coulomb correction factor $\CoulombFactor$ that
appears in \cref{eq:xsec1,eq:xsec2,eq:tot_xsec} is calculated using either the
Fermi function \cite{Fermi1934Original,Fermi1934Translation} or the modified
effective momentum approximation (MEMA) \cite{Engel1998}. Since the former is
known to overestimate Coulomb corrections at high lepton energies while the
latter does the same at low energies, the smaller of the two alternatives is
always chosen. This amounts to defining the Coulomb correction factor as
\begin{equation}
\label{eq:CoulombFactor}
\CoulombFactor \equiv \begin{cases}
  F_\text{Fermi} & |F_\text{Fermi} - 1| < |F_\text{MEMA} - 1| \\
  F_\text{MEMA} & \text{otherwise}
\end{cases}
\end{equation}
where the Fermi function is given by
\begin{align}
\nonumber
F_\text{Fermi} = \frac{2(1+S)}{\big[\Gamma(1+2S)\big]^2}
\, (2 \, & \gamma_{\text{rel}} \, \beta_{\text{rel}} \, m_\ell \, R)^{2S-2} \,
\\[0.5\baselineskip]
& e^{-\pi\,\eta} \, \left|\Gamma\left( S - i\eta \right) \right|^2
\label{eq:Fermi_function}
\end{align}
and
\begin{equation}
\label{eq:FMEMA}
F_\text{MEMA} \equiv \frac{ \FNRpLepEff \, \FNReLepEff }
{ \FNRpLep \, \FNReLep } \,.
\end{equation}

In \cref{eq:Fermi_function}
the quantity $S$ is defined in terms of the
fine structure constant $\alpha$ by
\begin{equation}
S \equiv \sqrt{1 - \alpha^2 Z_f^2\,}
\end{equation}
where $Z_f$ is the proton number of the
final nucleus.
The nuclear radius (in natural units) may be estimated as
\begin{equation}
\label{eq:nuclear_radius}
R \approx \frac{ 1.2 \, A^{1/3} \, \text{fm} }{ \hbar\,c }\,,
\end{equation}
and the Sommerfeld parameter
$\eta$ is given by
\begin{equation}
\label{eq:Sommerfeld_param}
\eta = \frac{ \alpha \, Z_f \, z_\ell }{ \beta_{\text{rel}} } \,.
\end{equation}
where $z_\ell$ is the electric charge (in units of the elementary charge) of
the final-state lepton.

Typical presentations of the correction factors defined in
\cref{eq:Fermi_function,eq:FMEMA}
neglect the small recoil
kinetic energy of the final nucleus in the laboratory frame. This allows the
use of expressions for $F_\text{Fermi}$ and $F_\text{MEMA}$ which are derived
in the rest frame of the final nucleus. I opt instead for Lorentz-invariant
forms of the correction factors written in terms of the relative speed
$\beta_{\text{rel}}$ of the two final-state particles: \cite{Cannoni2017}
\begin{align}
\label{eq:Lorentz_invariant_relative_speed}
\beta_{\text{rel}} &= \frac{ \sqrt{ (\plep \cdot \pNf)^2
- m_{\ell}^2 \, m_{f}^2 } }{ \plep \cdot \pNf }
&
\gamma_{\text{rel}} &\equiv \left(1 -
\beta^2_{\text{rel}}\right)^{-1/2}.
\end{align}
The symbols $\FNReLep$ and $\FNRpLep$ from \cref{eq:FMEMA}
denote, respectively, the total energy and momentum of the
outgoing lepton in the rest frame of recoiling nucleus:
\begin{align}
\FNReLep &\equiv \gamma_\text{rel} \, m_\ell
& \FNRpLep &\equiv \beta_\text{rel} \, \FNReLep\,.
\end{align}
The \textit{effective} values of these variables are those that exist in the
presence of the nuclear Coulomb potential
\begin{align}
\label{eq:eff_vars}
\FNRpLepEff &\equiv \sqrt{ \FNReLepEff^2 - m_\ell^2 } &
\FNReLepEff &\equiv \FNReLep - V_C(0),
\end{align}
which is approximated by that at the center of a uniformly-charged sphere:
\begin{equation}
V_C(0) \approx \frac{3 \, Z_f \, z_\ell \, \alpha}{2\,R}\,.
\end{equation}

It should be noted that, as originally presented \cite{Engel1998}, the MEMA
also involves modifying the value of the momentum transfer used to evaluate the
amplitude $\mathcal{M}$. Since the cross section treatment presented here
involves use of the long-wavelength limit $q \to 0$, however, I neglect this
additional correction.

\section{Allowed nuclear matrix elements}
\label{sec:allowed_MEs}

Despite sustained community interest and a concrete proposal by the CAPTAIN
experiment \cite{Berns2013} to perform direct measurements, no experimental
data for tens-of-\si{\MeV} charged-current $\nu_e$ scattering on argon are
currently available. However, in recent decades, three separate experiments
have obtained values of the allowed matrix elements $\FermiMat$ and $\GTMat$ by
considering related physics processes.

The first two experiments were performed in the late 1990s by separate teams
working at the Gesellschaft f\"{u}r Schwerionenforschung (Society for Heavy Ion
Research) in Darmstadt, Germany \cite{Liu1998} and the Grand Accélérateur
National d'Ions Lourds (Large Heavy Ion National Accelerator) in Caen, France
\cite{Bhattacharya1998}. Both sought to study CC $\nu_e$ absorption on
\isotope[40]{Ar} by measuring beta decays of its mirror nucleus
\isotope[40]{Ti}:
\begin{equation}
\isotope[40]{Ti} \rightarrow \isotope[40]{Sc}^{*} + e^{+} + \nu_e \,.
\end{equation}
In the limit of perfect isospin symmetry, the matrix element describing a
\isotope[40]{Ti} beta decay transition to a specific \isotope[40]{Sc} level is
equal to the matrix element accessing the level's isobaric analog in
\isotope[40]{K} via CC $\nu_e$ scattering on \isotope[40]{Ar}. The main
difficulties in applying this technique to neutrino cross sections are
\begin{enumerate*}[label=(\arabic*)]
\item the beta decay Q-value limits the maximum excitation energy that may be
studied, and
\item energy levels in the beta decay daughter nucleus (\isotope[40]{Sc}) must
be matched to their analogs in the final-state nucleus for neutrino scattering
(\isotope[40]{K}).
\end{enumerate*}

The third experiment \cite{Bhattacharya2009}, performed about a decade later at
the Indiana University Cyclotron Facility, extracted $\GTMat$ values from measurements of (p,n) scattering on
\isotope[40]{Ar}. The extraction technique relied on the observation, first put
forward in 1980 \cite{Goodman1980} and subsequently refined
\cite{Taddeucci1982, Taddeucci1987, Goodman2001}, that the (p,n) cross section
at very forward angles ($\theta \approx 0^\circ$) for
proton energies around \SI{100}{\MeV}
is approximately proportional to the allowed matrix elements
$\FermiMat$ and $\GTMat$. While subject to some unique difficulties of its own
(see, e.g., ref.~\cite[sec. 4.2]{Frekers2013}), this approach overcomes key
limitations of \isotope[40]{Ti} beta decay: transitions to excited levels of
\isotope[40]{K} may be studied directly at energies higher than the mirror beta
decay Q-value.

\subsection{Re-evaluation of existing measurements}
\label{sec:reeval}

Reasonable attempts were made in the original publications describing these
measurements to assign the extracted matrix elements to known \isotope[40]{K}
levels satisfying the spin-parity selection rules in
\crefrange{eq:Fermi_sel_rule}{eq:GT_sel_rule}. That is, the \isotope[40]{K}
isobaric analog state accessed via a Fermi transition must have $J^\pi =
0^{+}$, while GT transitions may only populate levels with $J^\pi =
1^{+}$. However, in light of new \isotope[40]{K} level data that became
available in 2017 \cite{Chen2017}, I revisited the level assignments for all
three measurements.

The results of this re-evaluation are shown in \cref{tab:evaluatedMEs}. Level
energies (\si{\keV}) and spin-parity assignments retrieved from the
Evaluated Nuclear Structure Data File (ENSDF)
database \cite{ENSDF} are listed in the first and second columns, respectively.
Excitation energies (for either \isotope[40]{Sc} or \isotope[40]{K} as
appropriate) and matrix element values are listed in the following columns for
each of the three experimental measurements. In the case of the (p,n)
scattering data, the matrix element values provided in the original paper
\cite{Bhattacharya2009} have been scaled by a factor of $g_A^2 = 1.26^2$. This
scaling was done because the definition of $\GTMat$ used by the experiment does
not include the axial-vector weak coupling constant $g_A$. The specific value
$g_A = 1.26$ was chosen for consistency with the one assumed in the
experimental analysis.

\begin{table*}
\centering
\caption{Level assignments and measured $\FermiMat$ and $\GTMat$ values
for
\isotope[\text{40}]{\text{Ar}}($\nu_e$,$e^{-}$)\isotope[\text{40}]{\text{K}^{*}}
}
\label{tab:evaluatedMEs}
\centering
\centerline{%
\begin{minipage}{0.98\textwidth}
\centering
\mysize
\renewcommand{\arraystretch}{1.3}
\renewcommand{\footnoterule}{\vspace{-8pt}}
\begin{tabular}{
@{\hspace{-6pt}} S[table-align-text-post=false] @{} c
@{\hspace{5pt}} S[table-alignment=right] @{\hspace{13pt}}
S[table-alignment=left, tight-spacing=true]
@{\hspace{10pt}}
S[table-alignment=right] @{\hspace{21pt}}
S[table-alignment=left, tight-spacing=true]
@{\hspace{15pt}}
S[table-alignment=right] @{\hspace{20pt}}
S[table-alignment=left, tight-spacing=true]
}
\toprule
{ \multirow{4}{2cm}{
\makecell[c]{
\\ Assigned \\ \isotope[\text{40}]{\text{K}} $E_x$ \\ \relax (\si{\keV}) \\ }}}
&
{ \multirow{4}{2.45cm}{
\makecell[c]{
ENSDF \cite{Chen2017}
\\ spin-parity \\ assignment\footnote{Parenthesized values are based upon
weak arguments \cite{Tuli2001}. } } } }
&
\multicolumn{2}{c}{{ \multirow{2}{2.7cm}{
\makecell[c]{
\footnotesize Liu \textit{et al.} \cite{Liu1998}
\\
\mysize \isotope[\text{40}]{\text{Ti}} $\beta^+$ decay}
}}}
&
\multicolumn{2}{c}{{ \multirow{2}{4.1cm}{
\makecell[c]{
\footnotesize Bhattacharya \textit{et al.} \cite{Bhattacharya1998}
\\
\mysize \isotope[\text{40}]{\text{Ti}} $\beta^+$ decay}
}}}
&
\multicolumn{2}{c}{{ \multirow{2}{4.1cm}{
\makecell[c]{
\footnotesize Bhattacharya \textit{et al.} \cite{Bhattacharya2009}
\\
\mysize
$\isotope[\text{40}]{\text{Ar}}(p,n)\isotope[\text{40}]{\text{K}}$ }
}}}
\\
&
& \multicolumn{2}{c}{ {} }
& \multicolumn{2}{c}{ {} }
& \multicolumn{2}{c}{ {} }
\\ \cmidrule(lr{8pt}){3-4}
\cmidrule(lr){5-6}
\cmidrule(lr){7-8}
& &
{ \multirow{2}{*}{
\makecell[c]{
\isotope[\text{40}]{\text{Sc}} $E_x$
\\ \relax (\si{\keV}) }}}
&
{ \multirow{2}{*}{
\makecell[c]{
$\FermiMat\, +$
\\ $\GTMat$ }}}
&
{ \multirow{2}{*}{
\makecell[c]{
\isotope[\text{40}]{\text{Sc}} $E_x$
\\ \relax (\si{\keV}) }}}
&
{ \multirow{2}{*}{
\makecell[c]{
$\FermiMat\, +$
\\ $\GTMat$ }}}
&
{ \multirow{2}{*}{
\makecell[c]{
\isotope[\text{40}]{\text{K}} $E_x$
\\ (\si{\keV}) }}}
&
{ \multirow{2}{*}{
\makecell[c]{
Weak\footnote{The
data tabulated in ref.~\cite{Bhattacharya2009}
were multiplied by $g_A^2 = 1.26^2$ to obtain
the $\GTMat$ values shown here.}
\\ $\GTMat$
}}}
\\ & & & & & & & \\
\midrule
2289.868(11) & 1$^{+}$ & 2287(10) & 0.83(8) & 2281(8) & 0.90(4) & 2333(30) &
1.64(16) \\
2730.357(19) & 1\hphantom{$^{+}$} & 2761(10) & 1.40(10) & 2752(8) & 1.50(6) &
2775(30) & 1.49(14) \\
2950.9(6) &  & 2966(40) & 0.03(1) & 2937(13) & 0.11(2) &  & \\
3109.56(4) & 1$^{+}$, 2$^+$ & 3121(46) & 0.06(3) & 3143(20) & 0.06(1) &  &  \\
3146.50(5) & 1$^{({-})}$ & 3235(50) & 0.16(4) &  &  & 3204(32) & 0.06(2) \\
3293(10) & unnatural\footnote{A nuclear level with parity $\Pi$ and spin $J$
has \textit{natural parity} if $\Pi = (-1)^J$. Otherwise it has
\textit{unnatural parity}. \label{foot:natural_parity} } & 3342(40) & 0.11(8)
& 3334(19) & 0.04(1) &  &  \\
3439.18(3) & (2$^{+}$) & 3418(60) & 0.05(2) &  &  &  &  \\
3517(15) &  & 3521(40) & 0.06(2) & 3569(56) & 0.01(1) & 3503(30) & 0.16(2) \\
3738.49(3)\footnote{Another candidate \isotope[\text{40}]{\text{K}} level for
this transition has $E_x = \SI{3663.88}{\keV}$ and $J^\pi = (1^{-}, 2, 3,
4^{+})$.} & 1$^{+}$ & 3662(40) & 0.13(7) & 3652(10) & 0.16(2) &  & \\
3797.48(3) & 1$^{+}$ & 3782(40) & 0.40(16) & 3786(10) & 0.26(3) &  & \\
3840.27(3) & (1, 2$^{+}$) &  &  & 3861(49) & 0.01(1) & 3870(30) & 0.44(5) \\
3996(10) & unnatural\cref{foot:natural_parity} & 4033(88) & 0.07(4) & 4067(24)
& 0.05(2) &  &  \\
4080(5) &  & 4194(60) & 0.10(6) & 4111(30) & 0.11(3) &  &  \\
4251.70(15) & (1, 2$^{-}$) & 4264(46) & 0.15(4) & 4267(10) & 0.29(3) &  &  \\
4383.7(7)\footnote{This level is the isobaric analog of the
\isotope[\text{40}]{\text{Ar}} ground state.} & 0$^{+}$ & 4365(10) & 4.01(31)
& 4364(8) & 3.84(17) &  & \\
4508(15) &  & 4540(86) & 0.14(5) & 4522(16) & 0.31(5) & 4421(30) & 0.86(14) \\
4697(10) & unnatural\cref{foot:natural_parity} & 4628(40) & 0.33(9) & 4655(12)
& 0.38(6) & & \\
4765(5) & (1)$^{+}$ & 4782(60) & 0.26(11) & 4825(21) & 0.47(8) & 4763(30) &
0.48(5) \\
4930(10) & unnatural\cref{foot:natural_parity} & 4997(72) & 0.24(10) &
5017(27) & 0.36(9) & & \\
5063.37(7) & (2$^{-}$, 3$^{+}$) & 5051(40) & 0.25(11) & 5080(35) & 0.23(7) &
& \\
5189.89(5) & (2$^{-}$) & 5135(86) & 0.20(6) & 5223(32) & 0.03(3) & 5162(30) &
0.59(6) \\
5247.1(6) &  & 5362(60) & 0.19(7) &  &  &  & \\
5488.65(17) & (2$^{-}$, 3, 4$^{-}$) & 5574(40) & 0.07(4) &  &  &  & \\
5681(32) &  & 5777(60) & 0.21(15) & 5696(23) & 0.11(4) & 5681(32) & 0.21(3) \\
5870(20) &  & 5886(80) & 0.17(7) &  &  &  &  \\
6118(30) &  & 6126(60) & 0.13(7) & 6006(21) & 0.13(5) & 6118(30) & 0.48(5) \\
6790(30) &  & 6426(60) & 0.11(6) &  &  & 6790(30) & 0.71(8) \\
7468(37) &  &  &  &  &  & 7468(37) & 0.06(2) \\
7795(33) &  &  &  &  &  & 7795(33) & 0.14(2) \\
7952(32) &  &  &  &  &  & 7952(32) & 0.97(10) \\
\midrule
{ Total\footnote{Gamow-Teller transitions are assumed for all levels
other than the isobaric analog state.} $B(\mathrm{GT})$ }
& & & 5.84(39) & & 5.52(20) & & 8.29(31) \\
\bottomrule
\end{tabular}
\label{tab:FandGT}
\end{minipage}
}
\end{table*}

\begin{figure}
\centering
\includegraphics{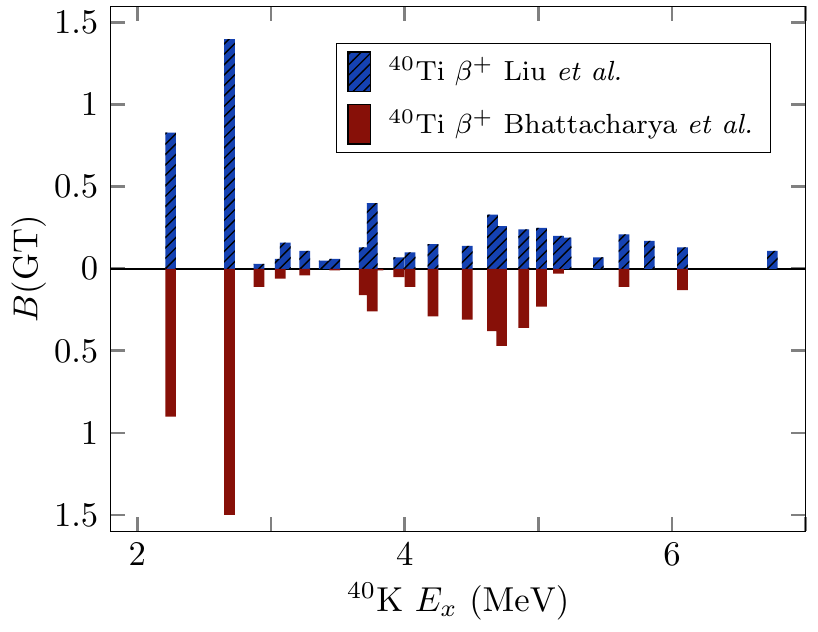}
\caption{Gamow-Teller strengths $\GTMat$ from two independent measurements of
\isotope[40]{Ti} $\beta^{+}$ decay by Liu \textit{et al.} \cite{Liu1998} and
Bhattacharya \textit{et al.} \cite{Bhattacharya1998}.}
\label{fig:BGT_comp_1}
\end{figure}

\begin{figure}
\centering
\includegraphics{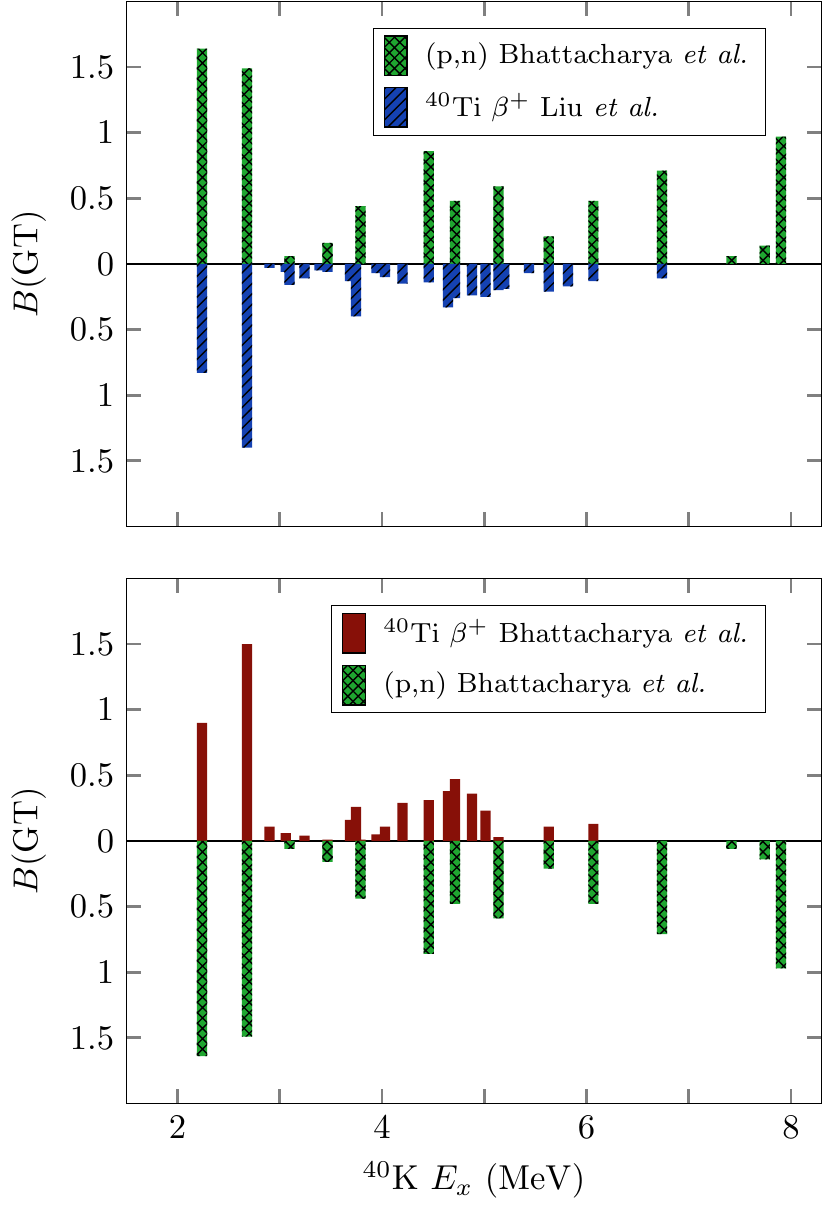}
\caption{Comparison of the Gamow-Teller strengths $\GTMat$ measured using
\isotope[40]{Ti} beta decay (see \cref{fig:BGT_comp_1}) with those obtained
using a measurement of $0^\circ$ (p,n) scattering
by Bhattacharya \textit{et al.} \cite{Bhattacharya2009}.}
\label{fig:BGT_comp_2}
\end{figure}

\Cref{fig:BGT_comp_1} shows the Gamow-Teller strengths obtained by the two
\isotope[40]{Ti} beta decay experiments mentioned previously. Excitation
energies of analog states in \isotope[40]{K}, represented on the horizontal
axis, are chosen to match the assignments made in \cref{tab:evaluatedMEs}. The
vertical axis is inverted for the second dataset to facilitate comparisons.
Rough consistency is seen between the two measurements, although the results
reported by Liu \textit{et al.} involve several more nuclear levels.

\Cref{fig:BGT_comp_2} uses a similar format to individually compare each beta
decay measurement to the GT strengths extracted from (p,n)
scattering. Substantially more fragmentation of the strength is seen in the
beta decay data, and there are areas of significant tension. For instance, the
two experimental methods disagree on whether the GT strength to the
\isotope[40]{K} level at \SI{2.3}{\MeV} is larger or smaller than the strength
to the level at \SI{2.7}{\MeV}.

Differences between the beta decay and (p,n) data were examined in detail by
Karako\c{c} \textit{et al.} in 2014 \cite{Karakoc2014}. Based on a combination
of theoretical calculations and a currently unpublished
$\isotope[40]{Ar}(\mathrm{h},\mathrm{t})\isotope[40]{K}$ measurement, the
authors argued that the (p,n) data should be preferred over the
\isotope[40]{Ti} beta decay data for calculations of CC $\nu_e$ absorption
cross sections on argon. Rather than attempt to adjudicate between the
conflicting datasets, I have opted to allow each of the three measurements to
be used as a source of $\GTMat$ values in \marley\ cross-section calculations.

\subsection{Extension to higher excitation energies}

Beyond the maximum excitation energy of about \SI{8}{\MeV} probed by the
experiments mentioned in the previous section, the presence of additional
Gamow-Teller strength is predicted by the model-independent Ikeda sum
rule~\cite{Ikeda1963}. This rule states that the summed GT strength
$S_\text{GT}^{-}$ ($S_\text{GT}^{+}$) over all possible nuclear final states
for CC $\nu_e$ ($\bar{\nu}_e$) absorption satisfies the relation
\begin{equation}
\label{eq:IkedaSumRule}
\Delta S_\text{GT} \equiv S_\text{GT}^{-} - S_\text{GT}^{+} = 3 \, g_A^2
\, (N_i - Z_i)
\end{equation}
where $N_i = 22$ ($Z_i = 18$) is the neutron (proton) number of the initial
\isotope[40]{Ar} nucleus. \Cref{eq:IkedaSumRule} implies a minimum possible
value of $S_\text{GT}^{-} = 12\,g_A^2 \approx 19$ for the integrated GT
strength associated with the reaction
$\isotope[40]{Ar}(\nu_e,e^{-})\isotope[40]{K}^{*}$. Comparing this value to the
measured total GT strengths listed in the final row of \cref{tab:evaluatedMEs}
reveals that the majority of the expected GT strength for CC $\nu_e$ absorption
on \isotope[40]{Ar} is unmeasured and associated with transitions to
high-lying, nucleon-unbound states of \isotope[40]{K}.

To supplement the experimental measurements with an estimate of the remainder
of the GT strength, I rely on a Quasiparticle Random Phase Approximation (QRPA)
calculation by Cheoun, Ha, and Kajino \cite{Cheoun2012a}. To avoid
double-counting, theoretical GT matrix elements are included with the
experimental ones only for excitation energies at which the integrated QRPA GT
strength exceeds the experimental total.

\begin{figure*}[t]
\centering
\includegraphics{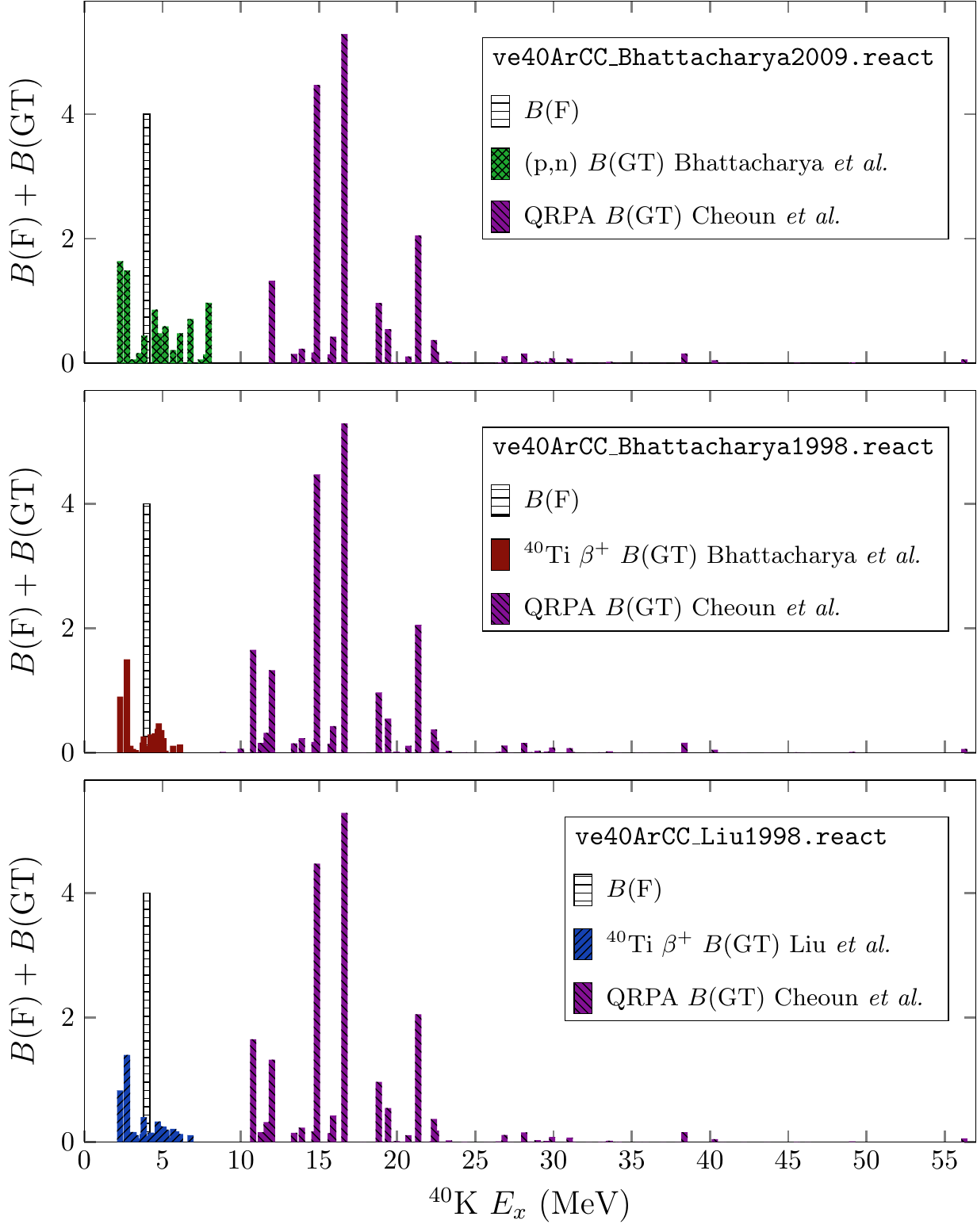}
\caption{Complete sets of $\isotope[40]{Ar}(\nu_e,e^{-})\isotope[40]{K}^*$
allowed approximation matrix elements distributed as part of \marley\ \version.
The name of the data file in which each set of matrix elements may be found is
given in the appropriate legend heading. The QRPA $B(\mathrm{GT})$ strengths
(violet) were calculated by Cheoun \textit{et al.} \cite{Cheoun2012a}.
Citations for the experimental $B(\mathrm{GT})$ strengths are given in
the captions for \cref{fig:BGT_comp_1,fig:BGT_comp_2}.}
\label{fig:marley_BF_plus_BGT}
\end{figure*}

\subsection{Adopted matrix elements}

\Cref{fig:marley_BF_plus_BGT} presents three complete sets of
allowed $\isotope[40]{Ar}(\nu_e,e^{-})\isotope[40]{K}^*$ nuclear
matrix elements prepared as input for \marley\ based on the measurements and
QRPA prediction discussed above. The experimental $\GTMat$ values are shown as
green, red, or blue bars depending on the dataset. The theoretical QRPA
$\GTMat$ values are shown in violet.

In addition to extracting GT strengths, both beta decay experiments measured
values of the Fermi matrix element $\FermiMat$. Under the approximation that
isospin is a good quantum number, this matrix element is expected to have the
value
\begin{equation}
\FermiMat = g_V^2 (N_i - Z_i) = 4
\end{equation}
and to connect the ground state of \isotope[40]{Ar} to a single $0^{+}$
isobaric analog state in \isotope[40]{K}, which has been identified as the
level with excitation energy $E_x = \SI{4.3837}{\MeV}$. Since the experimental
data are fully consistent with these expectations, I adopt the value $\FermiMat
= 4$ for this transition in all three sets of \marley\ matrix elements. The
known Fermi transition is represented in each panel of
\Cref{fig:marley_BF_plus_BGT} by a white bar with horizontal hatch marks.
Transitions to all other nuclear levels are assumed to proceed via the
Gamow-Teller operator.

Although the differences become important at neutrino energies near threshold,
the observables predicted in this paper are largely insensitive to the choice
between the three sets of \marley\ matrix elements for the
$\isotope[40]{Ar}(\nu_e,e^{-})\isotope[40]{K}^*$ reaction. For definiteness,
all \marley\ calculations shown in this work (see \cref{sec:results}) are
obtained using the \texttt{ve40ArCC\_Bhattacharya1998.react} input file, which
contains the matrix elements shown in the middle panel of
\cref{fig:marley_BF_plus_BGT}.

\section{Nuclear de-excitation model}
\label{sec:deex_model}

To model nuclear de-excitations following CC $\nu_e$ absorption on
\isotope[40]{Ar}, I rely on the observation that, due to the selection rules in
\cref{eq:Fermi_sel_rule,eq:GT_sel_rule}, each nuclear final state accessed by
the neutrino interaction has a well-defined excitation energy, spin, and
parity. Distinct treatments are used for bound and unbound nuclear states, with
the latter being those for which the excitation energy exceeds the separation
energy for at least one nuclear fragment with mass number $A \leq 4$.
Separation energies are computed in \marley\ using atomic and particle mass
data from refs.~\cite{CODATA2010,AME2012}. Untabulated atomic masses are
estimated using the liquid drop model of Myers and Swiatecki
\cite{Myers1966,Koning2008}. The full de-excitation cascade is treated
as a sequence of binary decays while neglecting the possibility of fission
and emission of heavy nucleon clusters ($A \geq 5$).

\subsection{Bound states: discrete level data}

De-excitations of bound nuclear states are handled in \marley\ using a set of
nuclear structure data files originally prepared for use with version 1.6 of
the TALYS \cite{Talys1,Talys2012,TALYS16} nuclear reaction code.\footnote{The
current release of TALYS is version 1.95. However, beginning with TALYS 1.8,
some nuclear levels for nuclei with an odd mass number (e.g., \isotope[39]{K})
appear in the structure data files with an unphysical integer spin (as opposed
to a half-integer spin). This bug, which remains unfixed \cite{TALYS195}, has
prevented updates to the data files distributed with \marley.} These in turn
are based on the level schemes included in version 3 of the Reference Input
Parameter Library (RIPL-3) \cite{RIPL3}. For a large number of nuclei, the
files provide tables of discrete nuclear energy levels including their
excitation energies, spin-parities, and de-excitation $\gamma$-ray branching
ratios. Missing experimental measurements of these quantities are supplemented
by theoretical estimates. Although internal conversion coefficients are
provided in the TALYS data files, this process is neglected in \marley\
\version. In the unusual case where discrete level data are not available for a
particular nuclide, $\gamma$-ray emission is simulated in the same manner as
for unbound nuclear states.

The TALYS structure data files are used with minor reformatting for all
nuclides except \isotope[40]{K}. To ensure consistency with the level
assignments made in \cref{sec:reeval}, I prepared an original decay scheme for
\isotope[40]{K} using the experimental data in ref.~\cite{Chen2017}, the TALYS
1.6 structure file, and (where needed) estimated $\gamma$-ray branching ratios
computed using the strength function defined in \cref{sec:gamma_tr_coeff}.

\subsection{Unbound states: statistical emission}
\label{sec:unbound}

The \marley\ approach to modeling de-excitations of unbound nuclear states
rests on the assumption of \textit{compound nucleus} formation: the neutrino
interaction leaves the nucleus in a state of thermal equilibrium which
de-excites independently of the manner in which it was formed. The number of
open decay channels is taken to be large enough that competition between them
can be modeled statistically. That is, transitions to individual nuclear final
states can be neglected in favor of averaging over many states of approximately
the same energy \cite{Weisskopf1937}. This last assumption is not strictly true
for excitation energies slightly above the lowest fragment emission threshold.
In such cases it is adopted as an approximation.

Compound nucleus modeling is a key ingredient in nuclear reaction codes
designed to compute nucleon-nucleus and nucleus-nucleus cross sections, such as
\mbox{TALYS}, \mbox{EMPIRE}~\cite{Empire2007}, \mbox{CCONE}~\cite{CCone2016},
and \mbox{CoH$_3$}~\cite{CoH3_2010}. The treatment in \marley\ uses the
Hauser-Feshbach formalism \cite{Hauser1952} common to these other codes.

\subsubsection{Differential decay widths}
\label{sec:diff_decay_widths}

For the present application
to neutrino-induced de-excitations, the physics content of the
Hauser-Feshbach statistical model (HFSM) may be conveniently summarized
by the differential decay widths
\begin{widetext}
\begin{align}
\label{eq:fragment_diff_decay_width}
\frac{ d\Gamma_{\fragment} }{ dE_x^\prime }
&= \frac{1}{2 \, \pi \, \rho_i(E_x, J, \Pi) }
\sum_{\ell = 0}^{ \infty }
\;
\sum_{j = |\ell - s|}^{\ell + s}
\;
\sum_{J^\prime = |J - j|}^{J + j}
T_{\ell j}(\totKinE) \, \rho_f(E_x^\prime, J^\prime, \Pi^\prime)
\intertext{and}
\label{eq:gamma_diff_decay_width}
\frac{ d\Gamma_\gamma }{ dE_x^\prime }
&= \frac{1}{2 \, \pi \, \rho_i(E_x, J, \Pi) }
\sum_{\lambda = 1}^{ \infty } \;
\sum_{J^\prime = |J - \lambda|}^{J + \lambda} \;
\sum_{ \Pi^\prime \in \{-1, 1\} }
T_{X\lambda}(E_\gamma)
\, \rho_f(E_x^\prime, J^\prime, \Pi^\prime)\,.
\end{align}
\end{widetext}
which describe de-excitations of a compound nuclear state via emission of a
fragment $\fragment$ or a $\gamma$-ray, respectively.

Here the initial (final) nucleus has excitation energy $E_x$ ($E_x^\prime$),
total spin $J$ ($J^\prime$), and parity $\Pi$ ($\Pi^\prime$); $s$, $\ell$, and
$j$ are the spin, orbital, and total angular momentum quantum numbers of the
emitted fragment; $\rho_i$ ($\rho_f$) is the density of nuclear levels in the
vicinity of the initial (final) state; $\totKinE$ is the total kinetic energy
of the final particles in the rest frame of the initial nucleus; and $E_\gamma$
is the energy of the emitted $\gamma$-ray. For decays involving emission of a
fragment with parity $\pi_\fragment$, the value of $\Pi^\prime$ is fixed by
conservation of parity:
\begin{equation}
\label{eq:HF_parity_conservation}
\Pi^\prime = (-1)^\ell \, \pi_\fragment \, \Pi \,.
\end{equation}
The possible $\gamma$-ray transitions are labeled by their multipolarity
$\lambda \geq 1$ and by whether they are \textit{electric} ($X = \mathrm{E}$)
or \textit{magnetic} ($X = \mathrm{M}$) in nature. These two alternatives are
distinguished based on the multipolarity and the nuclear parity:
\begin{equation}
X = \begin{cases}
\mathrm{E} & \Pi = (-1)^\lambda \, \Pi^\prime \\
\mathrm{M} & \Pi = (-1)^{\lambda + 1} \, \Pi^\prime \,.
\end{cases}
\end{equation}

The transmission coefficients $T_{\ell j}$ and $T_{X \lambda}$ quantify how
likely each decay mode is to occur. The methods used to compute them are
described in \cref{sec:fragment_tr_coeff,sec:gamma_tr_coeff}. For practical
calculations, the infinite sums that appear in
\cref{eq:fragment_diff_decay_width,eq:gamma_diff_decay_width} must be
truncated. Because the value of $T_{\ell j}$ ($T_{X \lambda}$) falls off
rapidly with increasing $\ell$ ($\lambda$), terms beyond $\ell = \lambda = 5$
are neglected.

The HFSM is often communicated in terms of nuclear scattering cross
sections instead of decay widths. To aid the reader in connecting the
expressions given here with more standard presentations (see, e.g.,
refs.~\cite{Cole2000,Thompson2009}), a brief derivation of
\cref{eq:fragment_diff_decay_width} is provided in \cref{sec:derive_width}.

\subsubsection{Fragment transmission coefficients}
\label{sec:fragment_tr_coeff}

The fragment transmission coefficients $T_{\ell j}$ used in
\cref{eq:fragment_diff_decay_width} are computed by solving the radial
Schr\"{o}dinger equation (with relativistic kinematics as recommended in
ref.~\cite{Koning2003})
\begin{equation}
\label{eq:rad_schrod}
\left[\frac{d^2}{dr^2} + \fragmentMomCM^2 - \frac{\ell(\ell + 1)}{r^2}
- \frac{\fragmentMomCM^2}{\totKinE}\,\mathcal{U}(r, \fragmentKinELab, \ell, j)
\right]u_{\ell j}(r) = 0
\end{equation}
where $u_{\ell j}$ is the fragment's radial wave function,
\begin{equation}
\label{eq:fragment_momentum_CM}
\fragmentMomCM = \sqrt{ \frac{ (2 \, m_\fragment + \fragmentKinELab)
\, {M^\prime}^{\,2} \, \fragmentKinELab } { (m_\fragment + M^{\prime})^2
+ 2 \, M^\prime \, \fragmentKinELab } }
\end{equation}
is the magnitude of its three-momentum in the rest frame of the initial nucleus,
and $m_\fragment$ ($M^\prime$) denotes the mass of the emitted fragment (final
nucleus). The global nuclear optical potential $\mathcal{U}$ developed by
Koning and Delaroche \cite{Koning2003} is used in the present calculations. A
full description thereof is given in \cref{sec:optical_potential}.

The quantity
\begin{equation}
\label{eq:fragment_kin_E_lab}
\fragmentKinELab = \frac{ \totKinE^2 + 2\, (m_\fragment + M^\prime)
\, \totKinE } { 2 M^\prime }
\end{equation}
is the fragment's kinetic energy in the rest frame of the final nucleus. The
label \textit{lab} is applied to this variable because it also represents the
laboratory-frame kinetic energy for the time-reversed process in which the
fragment is absorbed to form the compound nucleus (see
\cref{sec:derive_width}).

The transmission coefficient $T_{\ell j}$ is related to the energy-averaged
S-matrix element $\SMatrixElement$ via
\begin{equation}
T_{\ell j} = 1 - \left|\SMatrixElement\right|^{\,2} \,.
\end{equation}
The latter quantity may be determined by comparing the full solution
$u_{\ell j}$ of \cref{eq:rad_schrod} to the asymptotic form
\begin{equation}
\label{eq:asymptotic_radial_wavefunction}
u_{\ell j}(r) \to
\frac{i}{2}\left[H_\ell^-(\eta, \fragmentMomCM r)
- \SMatrixElement H_\ell^+(\eta, \fragmentMomCM r)\right]
\end{equation}
valid for large radii $r$, where the nuclear optical potential approaches the
Coulomb potential. Here $H_\ell^\pm$ are the Coulomb wave functions
\cite[chap.~33]{NIST:DLMF}. These depend on the Sommerfeld parameter
\begin{equation}
\eta \equiv \frac{z \, Z^\prime \,\alpha}{\beta_\text{rel}} \,,
\end{equation}
which is evaluated in terms of the proton number $z$ ($Z^\prime$) of the
emitted fragment (final nucleus) and the relative speed
\begin{equation}
\beta_\text{rel} = \frac{
\sqrt{ \fragmentKinELab^2 + 2 \, m_\fragment \, \fragmentKinELab  } }
{ m_\fragment + \fragmentKinELab } \,.
\end{equation}
The numerical techniques used to evaluate the fragment transmission
coefficients $T_{\ell j}$ are documented in ref.~\cite[sec.~2.2.2]{marleyCPC}.

\subsubsection{Gamma-ray transmission coefficients}
\label{sec:gamma_tr_coeff}

The $\gamma$-ray transmission coefficients $T_{X \lambda}$ used in
\cref{eq:gamma_diff_decay_width} may be written in terms of a strength function
$f_{X \lambda}(E_\gamma)$ such that
\begin{equation}
T_{X \lambda}(E_\gamma) = 2 \, \pi \, E_\gamma^{\,2\,\lambda + 1}
  \, f_{X \lambda}(E_\gamma) \,.
\end{equation}
To compute $\gamma$-ray strength functions in this work, I adopt the Standard
Lorentzian model from RIPL-3 \cite{RIPL3}, which is based on early studies by
Brink \cite{Brink1955} and Axel \cite{Axel1962}. This model assumes that
$\gamma$-ray emissions of type $X \lambda$ occur via de-excitation of the
corresponding giant multipole resonance, which is parameterized in terms of its
centroid excitation energy $E_{X \ell}$, width $\Gamma_{X \ell}$, and peak
cross section $\sigma_{X \ell}$. The strength function is evaluated according
to
\begin{equation}
\label{eq:SLM_strength_function}
f_{X \lambda}(E_\gamma) = \frac{ \sigma_{X \lambda} } { (2\,\lambda + 1)
\, \pi^2 } \bigg[ \frac{ \Gamma^2_{X \lambda} \, E_\gamma^{\,3 - 2\lambda} }{
(E_\gamma^{\,2} - E_{X \lambda}^2)^2 + E_\gamma^{\,2}
\, \Gamma^2_{X \lambda} }
\bigg]
\end{equation}
with the values of the giant resonance parameters given in
\cref{tab:giant_resonance_parameters}. Note that some peak cross sections are
given in the table in units of \si{\milli\barn} while
\cref{eq:SLM_strength_function} employs natural units.

\begin{table*}
\centering
\begin{minipage}{1\textwidth}
\centering
\begin{threeparttable}
\caption{Giant resonance parameters used herein for $\gamma$-ray strength
function calculations. Centroid excitation energies $E_{X\lambda}$ and widths
$\Gamma_{X\lambda}$ are given in \si{\MeV}. Peak cross sections for electric
multipole resonances ($\sigma_{E\lambda}$) are given in \si{\milli\barn}, while
those for magnetic resonances ($\sigma_{M\lambda}$) are given in
\si{\MeV\tothe{-2}}.}
\renewcommand{\arraystretch}{1.5}
\begin{tabular}{l@{\hspace{25pt}}l}
\toprule
Transition & Parameters \\
\midrule
Electric dipole (E1)\,\tnote{a}
& $E_\text{E1} = 31.2\,A^{-1/3} + 20.6\,A^{-1/6}$  \\
 & $\displaystyle\Gamma_\text{E1} = 0.026\,E_\text{E1}^{1.91}$ \\
 & $\sigma_\text{E1} = 1.2\,(120\,N\,Z)\,/\,(\pi \, A \, \Gamma_\text{E1})$ \\
\noalign{\vspace{4mm}}

Electric quadrupole (E2)\,\tnote{b}
& $E_\text{E2} = 63\,A^{-1/3}$  \\
 & $\displaystyle\Gamma_\text{E2} = 6.11 - 0.012\,A$ \\
 & $\sigma_\text{E2} = 0.00014\,Z^2 \, E_\text{E2}
 \, / \, (A^{1/3}\,\Gamma_\text{E2})$ \\
\noalign{\vspace{4mm}}

Magnetic dipole (M1)\,\tnote{c} \,\tnote{d}
& $E_\text{M1} = 41\,A^{-1/3}$  \\
 & $\displaystyle\Gamma_\text{M1} = 4$ \\
 & $\displaystyle\sigma_{\text{M}1}
 = 3\,\pi^2 \left[\frac{\left(B_\text{n}^2
 - E_{\text{M}1}^2\right)^2 + B_\text{n}^2\,
 \Gamma_{\text{M}1}^2}{B_\text{n}\,\Gamma_{\text{M}1}^2}
 \right]\Bigg[\frac{f_{\text{E}1}(B_\text{n})}{0.0588A^{0.878}}\Bigg]
 $ \\
\noalign{\vspace{4mm}}

Other electric transitions (E$\lambda$)\,\tnote{e}
& $E_{\text{E}\lambda} = E_\text{E2}$  \\
 & $\Gamma_{\text{E}\lambda} = \Gamma_\text{E2}$ \\
 & $\sigma_{\text{E}\lambda} = (\num{8e-4})^{\lambda - 2} \, \sigma_\text{E2}$ \\
\noalign{\vspace{4mm}}

\label{tab:giant_resonance_parameters}

Other magnetic transitions (M$\lambda$)\,\tnote{e}
& $E_{\text{M}\lambda} = E_\text{M1}$  \\
 & $\Gamma_{\text{M}\lambda} = \Gamma_\text{M1}$ \\
 & $\sigma_{\text{M}\lambda} = (\num{8e-4})^{\lambda - 1} \, \sigma_\text{M1}$ \\
\bottomrule
\end{tabular}
\begin{tablenotes}
\footnotesize
\item [a] See ref.~\cite[p. 129]{RIPL2}
\item [b] See ref.~\cite[p. 103]{RIPL1}
\item [c] See ref.~\cite[p. 132]{RIPL2}
\item [d] $B_\text{n}$ = \SI{7}{\MeV} and $f_{\text{E}1}$
 is calculated using natural units and the E1 parameters above.
\item [e] Default approximation used by version 1.6 of TALYS \cite{TALYS16}
\end{tablenotes}
\end{threeparttable}
\end{minipage}
\end{table*}

\subsubsection{Transitions to discrete nuclear levels}

The differential decay widths given in
\cref{eq:fragment_diff_decay_width,eq:gamma_diff_decay_width} are appropriate
for use at high excitation energies $E_x^\prime$ where the nuclear levels may
be modeled as a continuum. When a discrete level scheme is available for the
final-state nuclide, \marley\ uses the excitation energy of the last tabulated
level as the lower bound for the continuum. Otherwise, a continuum level
density $\rho_f$ is used all the way down to the ground state ($E_x^\prime =
0$).

Decays to discrete levels of the final-state nucleus are considered by \marley\
in terms of the HFSM partial decay widths
\begin{align}
\label{eq:fragment_partial_width}
\Gamma_\fragment &= \frac{1}{2 \, \pi \, \rho_i(E_x, J, \Pi) }
\sum_{j = |J - J^\prime|}^{J + J^\prime} \;
\sum_{\ell = |j - s|}^{j + s} \delta_\pi^\ell \,
T_{\ell j}(\totKinE)
\intertext{and}
\label{eq:gamma_partial_width}
\Gamma_\gamma &=
\frac{ 1 }{ 2 \, \pi \, \rho_i(E_x, J, \Pi) }
\sum_{\lambda = \max(1, |J - J^\prime|)}^{J + J^\prime}
T_{X\lambda}(E_\gamma)\,.
\end{align}
The symbol $\delta_\pi^\ell$, which enforces parity conservation, is equal to
one if \cref{eq:HF_parity_conservation} is satisfied and zero if it is not. If
$J+J^\prime < 1$, then the width $\Gamma_\gamma$ vanishes. The expressions in
\cref{eq:fragment_partial_width,eq:gamma_partial_width} may be derived from
\cref{eq:fragment_diff_decay_width,eq:gamma_diff_decay_width} by treating
$\rho_f$ as a delta function centered on the nuclear level of interest.

\subsubsection{Nuclear level density}

In the continuum, the final level density $\rho_f$ is computed according to the
Back-shifted Fermi gas model (BFM) described in \cref{sec:level_density_model}.
The initial level density $\rho_i$ is evaluated according to the BFM at all
excitation energies. However, since the overall factor involving $\rho_i$
cancels out in the evaluation of decay branching ratios, the specific model
chosen for $\rho_i$ does not have any impact on the simulation results.

\section{Results}
\label{sec:results}

In this section, the \marley\ \version\ implementation of the theoretical
models described above is used to obtain predictions of total and differential
cross sections for the reaction
$\isotope[40]{Ar}(\nu_e,e^{-})\isotope[40]{K}^*$. Because \marley\ calculates
the four-momentum of every final-state particle for every event, various
additional distributions may be studied beyond those presented in this work.

\subsection{Inclusive cross section}

\begin{figure}
\includegraphics[width=\columnwidth]{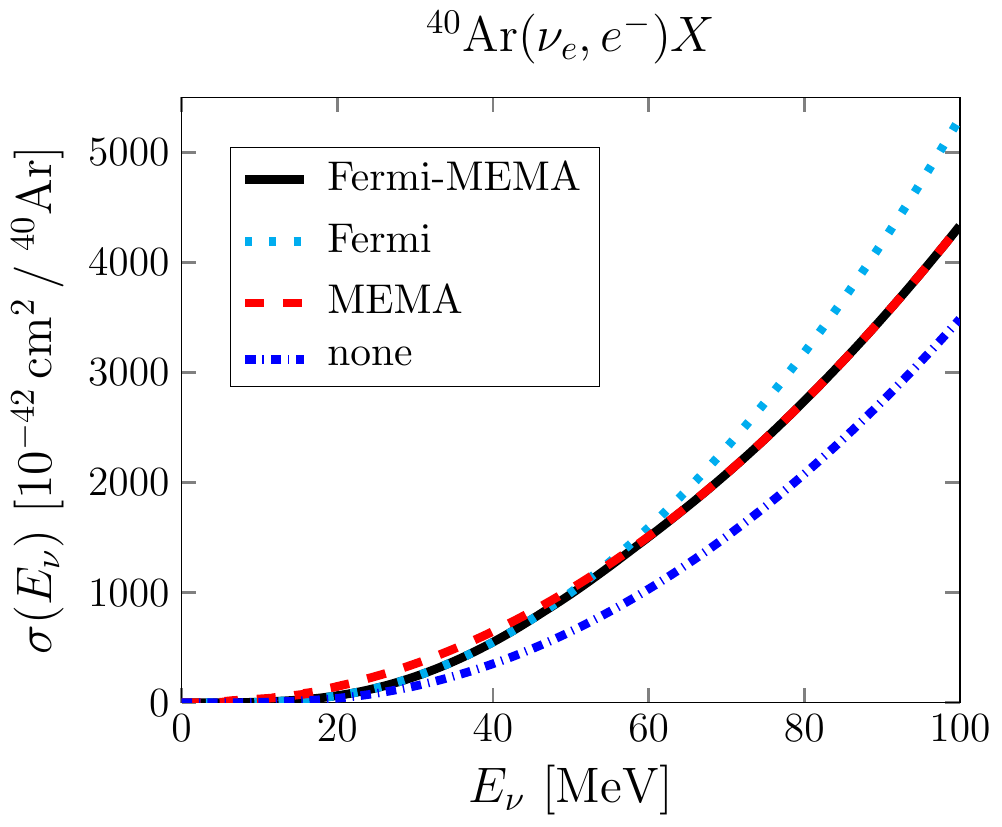}
\caption{Inclusive total cross sections for charged-current absorption
of $\nu_e$ on \isotope[40]{Ar}. Each curve shows the result obtained
using a specific approach to Coulomb corrections.}
\label{fig:total_xsec_coulomb}
\end{figure}

\Cref{fig:total_xsec_coulomb} shows \marley\ predictions of the total cross
section for inclusive charged-current $\nu_e$ absorption on \isotope[40]{Ar}.
The important role played by the Coulomb corrections discussed in
\cref{sec:Coulomb} is illustrated by the different curves in the plot. The
default \marley\ approach to Coulomb effects, defined in
\cref{eq:CoulombFactor}, involves choosing the smaller of two correction
factors calculated using the Fermi function and using the modified effective
momentum approximation (MEMA). In \cref{fig:total_xsec_coulomb}, the solid
black line gives the cross section calculated using the default approach, while
the dotted cyan and dashed red lines give, respectively, the corresponding
cross sections obtained when the Fermi function and MEMA are used
unconditionally. Applying either correction leads to an enhancement of the
total cross section over the uncorrected result, which is drawn as the blue
dash-dotted line. The relationships between the different approaches to Coulomb
corrections in the present calculation are qualitatively similar to those seen
previously using a CRPA model \cite{CRPA1}, but there are some details that
are different, e.g., the cross sections calculated using the Fermi function and
the MEMA intersect at a neutrino energy between
\SIrange[range-phrase=--]{50}{60}{\MeV}, about \SI{10}{\MeV} lower than in the
CRPA result.

\subsection{Exclusive cross sections}

\begin{figure}
\includegraphics[width=\columnwidth]{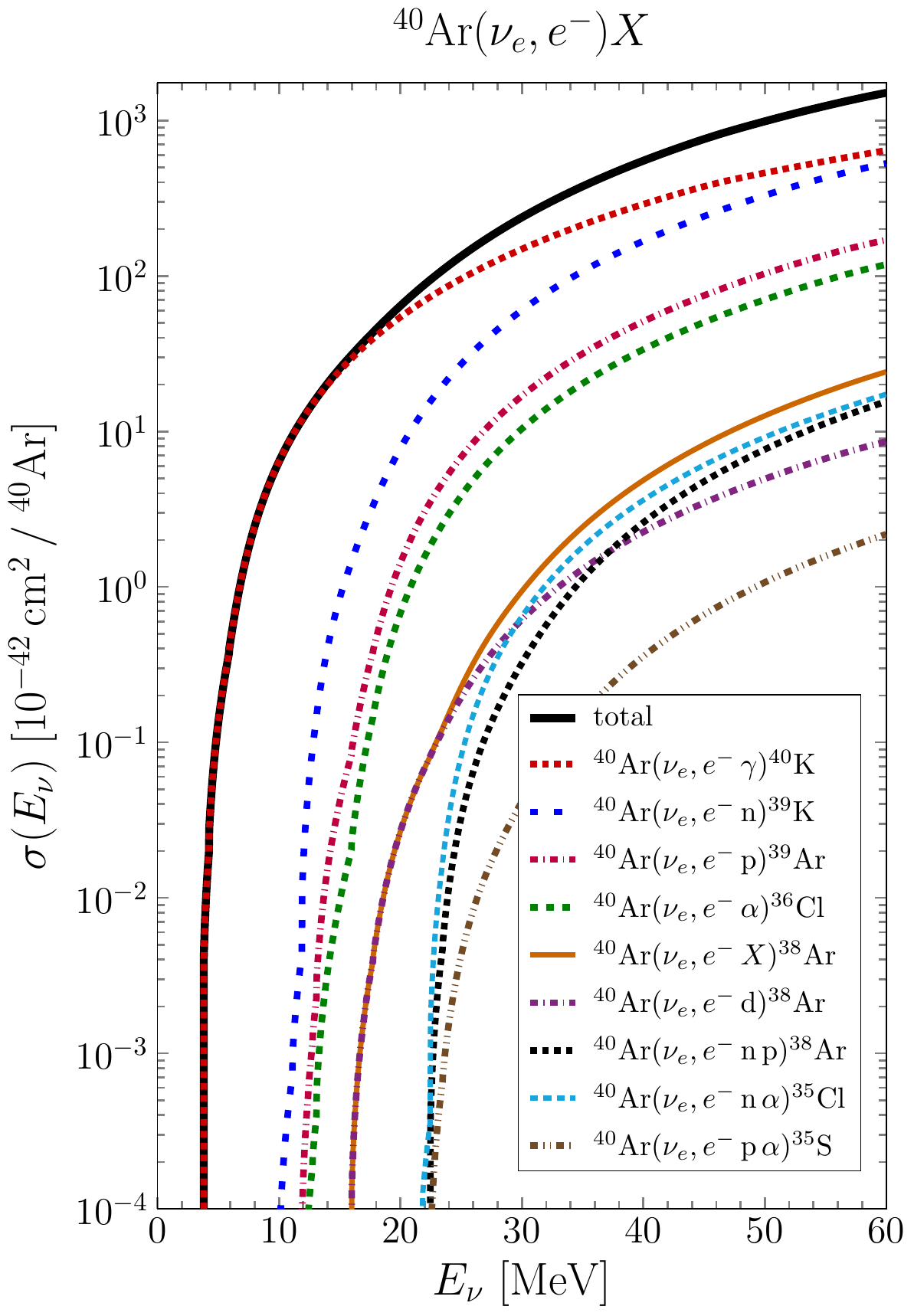}
\caption{Exclusive total cross sections for charged-current absorption
of $\nu_e$ on \isotope[40]{Ar}.}
\label{fig:exclusive_xsecs}
\end{figure}

\Cref{fig:exclusive_xsecs} presents the first calculation at supernova energies
of total cross sections for exclusive final states in the reaction
$\isotope[40]{Ar}(\nu_e,e^{-})\isotope[40]{K}^{*}$. Each exclusive channel is
labeled in terms of its hadronic content, but zero or more de-excitation
$\gamma$-rays are allowed even when not explicitly listed. Below a neutrino
energy of about \SI{10}{\MeV}, only transitions to bound nuclear levels are
energetically possible. These de-excite via $\gamma$-ray emission. Single
neutron emission becomes appreciable around \SI{15}{\MeV}. Although the proton
(\SI{7.58}{\MeV}) and alpha particle (\SI{6.44}{\MeV}) separation energies for
\isotope[40]{K} are comparable to the neutron separation energy
(\SI{7.80}{\MeV}), the Coulomb barrier experienced by these charged particles
suppresses their emission relative to neutrons.

Throughout the remainder of this paper, calculations of flux-averaged cross
sections will be reported for two distinct sources of low-energy electron
neutrinos. The first of these is the approximate supernova neutrino energy
spectrum
\begin{equation}
\phi(E_\nu) \propto \left( \frac{ E_\nu }{ \Eavg } \right)^{\alpha}
\exp\left[ -(\alpha + 1) \, \frac{ E_\nu }{ \Eavg } \right]
\end{equation}
described in ref.~\cite{Nikrant2018}. Here the dependence on the neutrino
energy $E_\nu$ is expressed in terms of the mean energy $\Eavg$ and a shape
parameter $\alpha$. Based on an analysis of a simulated supernova, the authors
of ref.~\cite{Nikrant2018} report values of $\Eavg = \SI{14.1}{\MeV}$ and
$\alpha = 2.67$ for the time-integrated $\nu_e$ spectrum, which I denote by
SN$_T$. I also consider four instantaneous spectra estimated using fig.~1 of
ref.~\cite{Nikrant2018}. These are labeled SN$_1$ through SN$_4$ in
chronological order. \Cref{tab:SN_params} gives the values of the spectral
parameters and the approximate elapsed time since the start of the supernova
for each of these configurations.

\begin{table}
\caption{Parameters used to compute the supernova neutrino spectrum described
in the text. Values of the mean $\nu_e$ energy $\Eavg$ and the
elapsed time $t$ are given in \si{\MeV} and seconds, respectively. The
shape parameter $\alpha$ is dimensionless.}
\label{tab:SN_params}
\begin{tabular}{l@{\hspace{9pt}}c@{\hspace{9pt}}c@{\hspace{9pt}}c}
\toprule
Configuration & $\Eavg$ & $\alpha$ & $t$ \\
\midrule
SN$_T$  & 14.1 & 2.67 & \\
SN$_1$ & 9 & 5 & 0 \\
SN$_2$ & 12 & 4 & 0.5 \\
SN$_3$ & 15 & 3.25 & 3 \\
SN$_4$ & 17 & 2.9 & 6.25 \\
\bottomrule
\end{tabular}
\end{table}

The second source of low-energy electron neutrinos considered in this work
is the decay
\begin{equation}
\mu^{+} \rightarrow e^{+} + \nu_e + \bar{\nu}_\mu\,.
\end{equation}
For an antimuon decaying at rest ($\mu$DAR), the $\nu_e$ energy spectrum is given by
\cite{Kosmas2017}
\begin{equation}
\phi(E_\nu) \propto E_\nu^2 \, m_\mu^{-4} \, (m_\mu - 2E_\nu) \,,
\end{equation}
where $m_\mu$ is the muon mass
and the neutrino energy $E_\nu$ satisfies
\begin{equation}
0 < E_\nu < m_\mu/2 \,.
\end{equation}
Experimental facilities which generate large numbers of stopped muons, such as
the Spallation Neutron Source at Oak Ridge National Laboratory, provide a
valuable opportunity to study tens-of-\si{\MeV} neutrino interactions using a
terrestrial source \cite{Bolozdynya2012}.

\Cref{tab:flux_avg_xsec_table} reports a wide variety of flux-averaged total
cross sections for each of the electron neutrino spectra $\phi(E_\nu)$
described above. For each entry in the table, the flux-averaged total cross
section $\FluxAvgTotXSec_f$ for a specific final state $f$ was obtained via the
expression
\begin{equation}
\label{eq:flux_avg_xsec1}
\FluxAvgTotXSec_f = \frac{ 1 }{ \Phi } \, \int \phi(E_\nu)
\left[ \sum_L \sigma_L(E_\nu) \, R_L(f) \right] dE_\nu
\end{equation}
where
\begin{equation}
\label{eq:flux_avg_xsec2}
\Phi \equiv \int \phi(E_\nu) \, dE_\nu
\end{equation}
and the integrals in \cref{eq:flux_avg_xsec1,eq:flux_avg_xsec2} are over the
entire neutrino spectrum. Here $\sigma_L$ is the inclusive total cross section
for transitions to a particular \isotope[40]{K} nuclear level $L$ and $R_L(f)$
is the branching ratio for the final state $f$ when the de-excitation cascade
begins at the level $L$. The sum in \cref{eq:flux_avg_xsec1} runs over all
energetically-accessible nuclear levels. All quantities in
\cref{eq:flux_avg_xsec1} are calculated analytically except for $R_L(f)$, which
is estimated using Monte Carlo simulations of de-excitations from every nuclear
level listed in the \marley\ input file
\texttt{ve40ArCC\_Bhattacharya1998.react}. The statistical uncertainty
associated with each entry in \mbox{\cref{tab:flux_avg_xsec_table}} never
exceeds 10\% and is typically much smaller.

\begin{table*}
\sisetup{table-format = +1.1e+2,
retain-zero-exponent=true,
}
\setlength{\extrarowheight}{0.25em}
\centering
\caption{Flux-averaged total cross sections computed for several different $\nu_e$ spectra
described in the text. All numerical values are given in $10^{-42}$
\si{\centi\meter\squared} $ / $ \isotope[40]{Ar}. For example,
a table entry of $4.5\times10^{1}$ should be interpreted as
$4.5\times10^{-41} \, \si{\centi\meter\squared} \, / \, \isotope[40]{Ar}$.}
\label{tab:flux_avg_xsec_table}
\begin{tabular}{l@{\hskip 1mm}S[table-auto-round]@{\hskip 1mm}S[table-auto-round]@{\hskip 1mm}S[table-auto-round]@{\hskip 1mm}S[table-auto-round]@{\hskip 1mm}S[table-auto-round]@{\hskip 1mm}S[table-auto-round]}
\toprule
Channel & {SN$_T$} & {$\mu$DAR} & {SN$_1$} & {SN$_2$} & {SN$_3$} & {SN$_4$} \\
\midrule
$\isotope[40]{Ar}(\nu_e, e^{-})X$   &  4.485471e+01 &  3.561866e+02 &  7.722795e+00 &  2.248090e+01 &  5.015778e+01 &  7.655412e+01 \\
$\isotope[40]{Ar}(\nu_e, e^{-} \, \gamma)\isotope[40]{K}$   &  3.292455e+01 &  1.941676e+02 &  7.335443e+00 &  1.896931e+01 &  3.678237e+01 &  5.176244e+01 \\
$\isotope[40]{Ar}(\nu_e, e^{-} \, \mathrm{n})\isotope[39]{K}$   &  8.221183e+00 &  1.044471e+02 &  3.128591e-01 &  2.570796e+00 &  9.232876e+00 &  1.669965e+01 \\
$\isotope[40]{Ar}(\nu_e, e^{-} \, \mathrm{p})\isotope[39]{Ar}$   &  2.161494e+00 &  3.140200e+01 &  4.918244e-02 &  5.786560e-01 &  2.420270e+00 &  4.613263e+00 \\
$\isotope[40]{Ar}(\nu_e, e^{-} \, \mathrm{d})\isotope[38]{Ar}$   &  7.752669e-02 &  1.372191e+00 &  1.018324e-03 &  1.699856e-02 &  8.613345e-02 &  1.773532e-01 \\
$\isotope[40]{Ar}(\nu_e, e^{-} \, \mathrm{t})\isotope[37]{Ar}$   &  9.703949e-04 &  2.316395e-02 &  3.368466e-06 &  1.417373e-04 &  1.060080e-03 &  2.491569e-03 \\
$\isotope[40]{Ar}(\nu_e, e^{-} \, \mathrm{h})\isotope[37]{Cl}$   &  7.203451e-05 &  1.869776e-03 &  2.031919e-07 &  9.583260e-06 &  7.818488e-05 &  1.912196e-04 \\
$\isotope[40]{Ar}(\nu_e, e^{-} \, \alpha)\isotope[36]{Cl}$   &  1.310415e+00 &  2.062336e+01 &  2.383907e-02 &  3.240405e-01 &  1.463157e+00 &  2.875723e+00 \\
$\isotope[40]{Ar}(\nu_e, e^{-} \, \mathrm{n} \, \mathrm{p})\isotope[38]{Ar}$   &  5.801653e-02 &  1.617583e+00 &  1.371793e-04 &  7.082974e-03 &  6.257303e-02 &  1.588299e-01 \\
$\isotope[40]{Ar}(\nu_e, e^{-} \, \mathrm{n}\,\mathrm{d})\isotope[37]{Cl}$   &  4.098699e-05 &  1.710917e-03 &  6.910478e-09 &  1.922290e-06 &  4.142292e-05 &  1.454654e-04 \\
$\isotope[40]{Ar}(\nu_e, e^{-} \, \mathrm{n} \, \alpha)\isotope[35]{Cl}$   &  8.954764e-02 &  2.187162e+00 &  2.938477e-04 &  1.275645e-02 &  9.765275e-02 &  2.320456e-01 \\
$\isotope[40]{Ar}(\nu_e, e^{-} \, \mathrm{p} \, \alpha)\isotope[35]{S}$   &  7.864443e-03 &  2.232536e-01 &  1.783766e-05 &  9.391102e-04 &  8.469125e-03 &  2.168641e-02 \\
$\isotope[40]{Ar}(\nu_e, e^{-} \, 2\mathrm{n} \, \mathrm{p})\isotope[37]{Ar}$   &  1.547900e-04 &  6.712654e-03 &  2.659833e-09 &  3.782569e-06 &  1.500282e-04 &  6.260851e-04 \\
$\isotope[40]{Ar}(\nu_e, e^{-} \, 2\alpha)\isotope[32]{Si}$   &  6.369740e-04 &  2.349068e-02 &  3.360093e-07 &  4.680675e-05 &  6.666756e-04 &  1.988936e-03 \\
$\isotope[40]{Ar}(\nu_e, e^{-} \, 2\mathrm{n})\isotope[38]{K}$   &  9.181615e-04 &  3.614841e-02 &  2.800241e-07 &  5.770515e-05 &  9.516680e-04 &  2.976387e-03 \\
$\isotope[40]{Ar}(\nu_e, e^{-} \, \mathrm{n} \, 2\mathrm{p})\isotope[37]{Cl}$   &  2.095507e-04 &  9.116601e-03 &  4.588246e-09 &  5.372580e-06 &  2.036794e-04 &  8.406027e-04 \\
$\isotope[40]{Ar}(\nu_e, e^{-} \, 2\mathrm{p})\isotope[38]{Cl}$   &  6.404862e-04 &  2.376367e-02 &  3.315336e-07 &  4.643078e-05 &  6.695140e-04 &  2.009746e-03 \\
$\isotope[40]{Ar}(\nu_e, e^{-} \, 3\mathrm{n})\isotope[37]{K}$   &  4.154733e-06 &  1.772791e-04 &  6.181527e-11 &  9.776563e-08 &  4.015836e-06 &  1.694344e-05 \\
$\isotope[40]{Ar}(\nu_e, e^{-} \, 3\mathrm{p})\isotope[37]{S}$   &  3.591039e-07 &  1.460174e-05 &  4.681105e-12 &  8.012472e-09 &  3.454761e-07 &  1.485352e-06 \\
$\isotope[40]{Ar}(\nu_e, e^{-} \, X)\isotope[28]{Al}$   &  2.235610e-08 &  6.552005e-07 &  1.844574e-13 &  4.013386e-10 &  2.107268e-08 &  9.823915e-08 \\
$\isotope[40]{Ar}(\nu_e, e^{-} \, X)\isotope[30]{P}$   &  2.569873e-09 &  2.930011e-09 &  1.097592e-15 &  1.966075e-11 &  2.267468e-09 &  1.346372e-08 \\
$\isotope[40]{Ar}(\nu_e, e^{-} \, X)\isotope[30]{Si}$   &  1.539868e-07 &  3.154833e-06 &  8.630280e-13 &  2.300615e-09 &  1.427498e-07 &  7.093741e-07 \\
$\isotope[40]{Ar}(\nu_e, e^{-} \, X)\isotope[31]{P}$   &  4.332934e-05 &  1.881430e-03 &  8.678348e-10 &  1.089998e-06 &  4.206660e-05 &  1.744072e-04 \\
$\isotope[40]{Ar}(\nu_e, e^{-} \, X)\isotope[31]{Si}$   &  3.066393e-06 &  1.301006e-04 &  4.608830e-11 &  7.207959e-08 &  2.963083e-06 &  1.251621e-05 \\
$\isotope[40]{Ar}(\nu_e, e^{-} \, X)\isotope[32]{P}$   &  6.369792e-04 &  2.349069e-02 &  3.360093e-07 &  4.680679e-05 &  6.666801e-04 &  1.988962e-03 \\
$\isotope[40]{Ar}(\nu_e, e^{-} \, X)\isotope[33]{P}$   &  2.602884e-07 &  8.538772e-06 &  3.053539e-12 &  5.213132e-09 &  2.471098e-07 &  1.121934e-06 \\
$\isotope[40]{Ar}(\nu_e, e^{-} \, X)\isotope[33]{S}$   &  2.822196e-06 &  1.015368e-04 &  5.948095e-11 &  6.504966e-08 &  2.707282e-06 &  1.180065e-05 \\
$\isotope[40]{Ar}(\nu_e, e^{-} \, X)\isotope[34]{Cl}$   &  2.756524e-05 &  1.182512e-03 &  4.249723e-10 &  6.552929e-07 &  2.666446e-05 &  1.121527e-04 \\
$\isotope[40]{Ar}(\nu_e, e^{-} \, X)\isotope[34]{P}$   &  1.359927e-06 &  5.610901e-05 &  1.885264e-11 &  3.100535e-08 &  1.310547e-06 &  5.596867e-06 \\
$\isotope[40]{Ar}(\nu_e, e^{-} \, X)\isotope[34]{S}$   &  3.416978e-04 &  1.489999e-02 &  1.214284e-08 &  9.619152e-06 &  3.339035e-04 &  1.349221e-03 \\
$\isotope[40]{Ar}(\nu_e, e^{-} \, X)\isotope[35]{Cl}$   &  8.954772e-02 &  2.187163e+00 &  2.938477e-04 &  1.275645e-02 &  9.765283e-02 &  2.320460e-01 \\
$\isotope[40]{Ar}(\nu_e, e^{-} \, X)\isotope[35]{S}$   &  7.864456e-03 &  2.232537e-01 &  1.783766e-05 &  9.391103e-04 &  8.469137e-03 &  2.168648e-02 \\
$\isotope[40]{Ar}(\nu_e, e^{-} \, X)\isotope[36]{Ar}$   &  1.306030e-05 &  5.298376e-04 &  6.763852e-10 &  3.891657e-07 &  1.276353e-05 &  5.163422e-05 \\
$\isotope[40]{Ar}(\nu_e, e^{-} \, X)\isotope[36]{Cl}$   &  1.310425e+00 &  2.062375e+01 &  2.383908e-02 &  3.240407e-01 &  1.463167e+00 &  2.875764e+00 \\
$\isotope[40]{Ar}(\nu_e, e^{-} \, X)\isotope[36]{S}$   &  5.319084e-07 &  2.025986e-05 &  9.538410e-12 &  1.222793e-08 &  5.112601e-07 &  2.209211e-06 \\
$\isotope[40]{Ar}(\nu_e, e^{-} \, X)\isotope[37]{Ar}$   &  1.166172e-03 &  3.158752e-02 &  3.378036e-06 &  1.474422e-04 &  1.251531e-03 &  3.263119e-03 \\
$\isotope[40]{Ar}(\nu_e, e^{-} \, X)\isotope[37]{Cl}$   &  3.034012e-04 &  1.189716e-02 &  2.113727e-07 &  1.594868e-05 &  3.038462e-04 &  1.110044e-03 \\
$\isotope[40]{Ar}(\nu_e, e^{-} \, X)\isotope[38]{Ar}$   &  1.355432e-01 &  2.989774e+00 &  1.155503e-03 &  2.408154e-02 &  1.487065e-01 &  3.361831e-01 \\
\bottomrule
\end{tabular}
\end{table*}

\subsection{Electron angle and energy distributions}

\begin{figure}
\includegraphics[width=0.99\columnwidth]{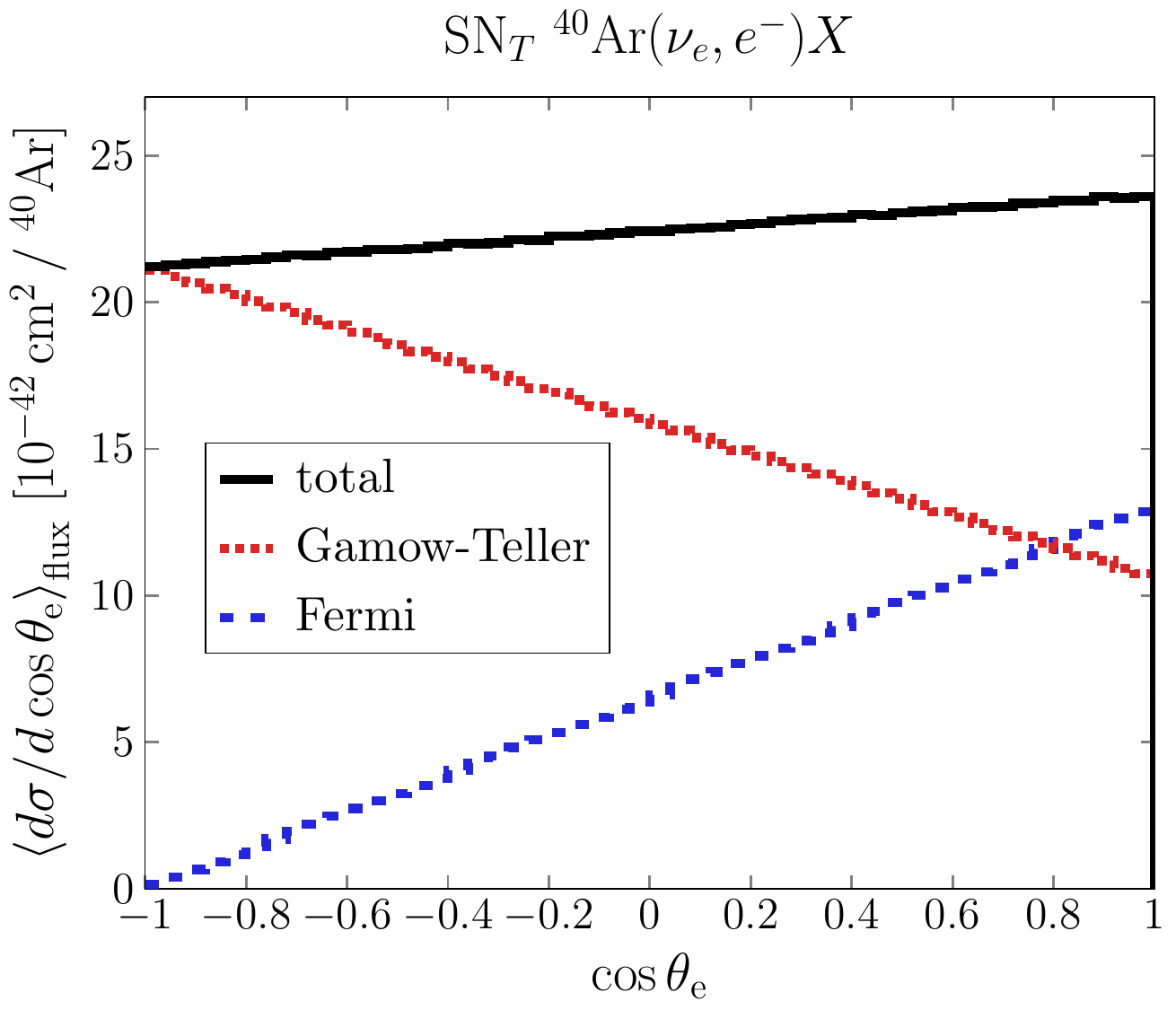}
\\[\baselineskip]
\includegraphics[width=0.99\columnwidth]{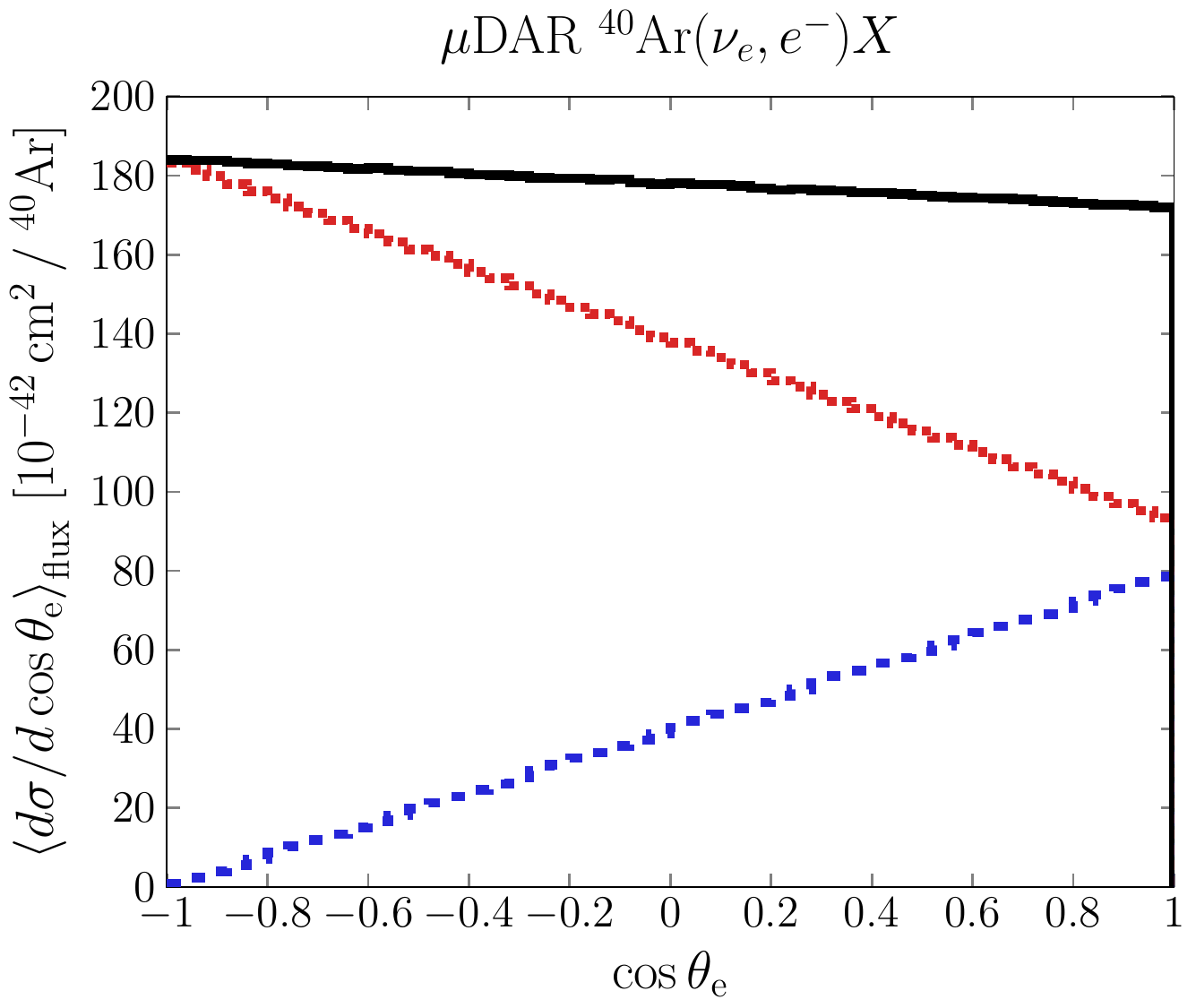}
\caption{Flux-averaged differential cross sections with respect to the
laboratory-frame scattering cosine $\cos\theta_\mathrm{e}$ of the final-state
electron. Top: Calculation for the time-integrated supernova $\nu_e$ spectrum
described in the text. Bottom: Calculation for $\nu_e$ produced by $\mu^{+}$
decay at rest.}
\label{fig:diff_xsecs_3}
\end{figure}

\Cref{fig:diff_xsecs_3} shows flux-averaged differential cross sections
predicted by \marley\ for the laboratory-frame scattering cosine of the outgoing
electron. The upper panel shows the total result (solid black) for the SN$_T$
$\nu_e$ spectrum together with the separate contributions arising from Fermi
(dashed blue) and Gamow-Teller (dotted red) transitions. The lower panel
presents the same quantities for the $\mu$DAR $\nu_e$ spectrum. Competition
between the two linear components of the cross section gives rise to a total
angular distribution that is nearly flat in both cases, with SN$_T$ being
slightly forward and $\mu$DAR slightly backward.

A recent theoretical study \cite{CRPA2} has pointed out that forbidden nuclear
transitions, which are neglected in the present calculation, have an
increasingly strong effect on the electron angular distribution as the neutrino
energy grows beyond a few tens of \si{\MeV}. Deviations from the linear
behavior shown in \cref{fig:diff_xsecs_3} signal the breakdown of the allowed
approximation used by \marley. A future measurement of the
$\isotope[40]{Ar}(\nu_e,e^{-})\isotope[40]{K}^{*}$ angular differential cross
section will thus provide a powerful constraint on the nuclear modeling needed
to predict the relative contributions of the allowed and forbidden transitions.

\begin{figure}
\includegraphics[width=0.99\columnwidth]{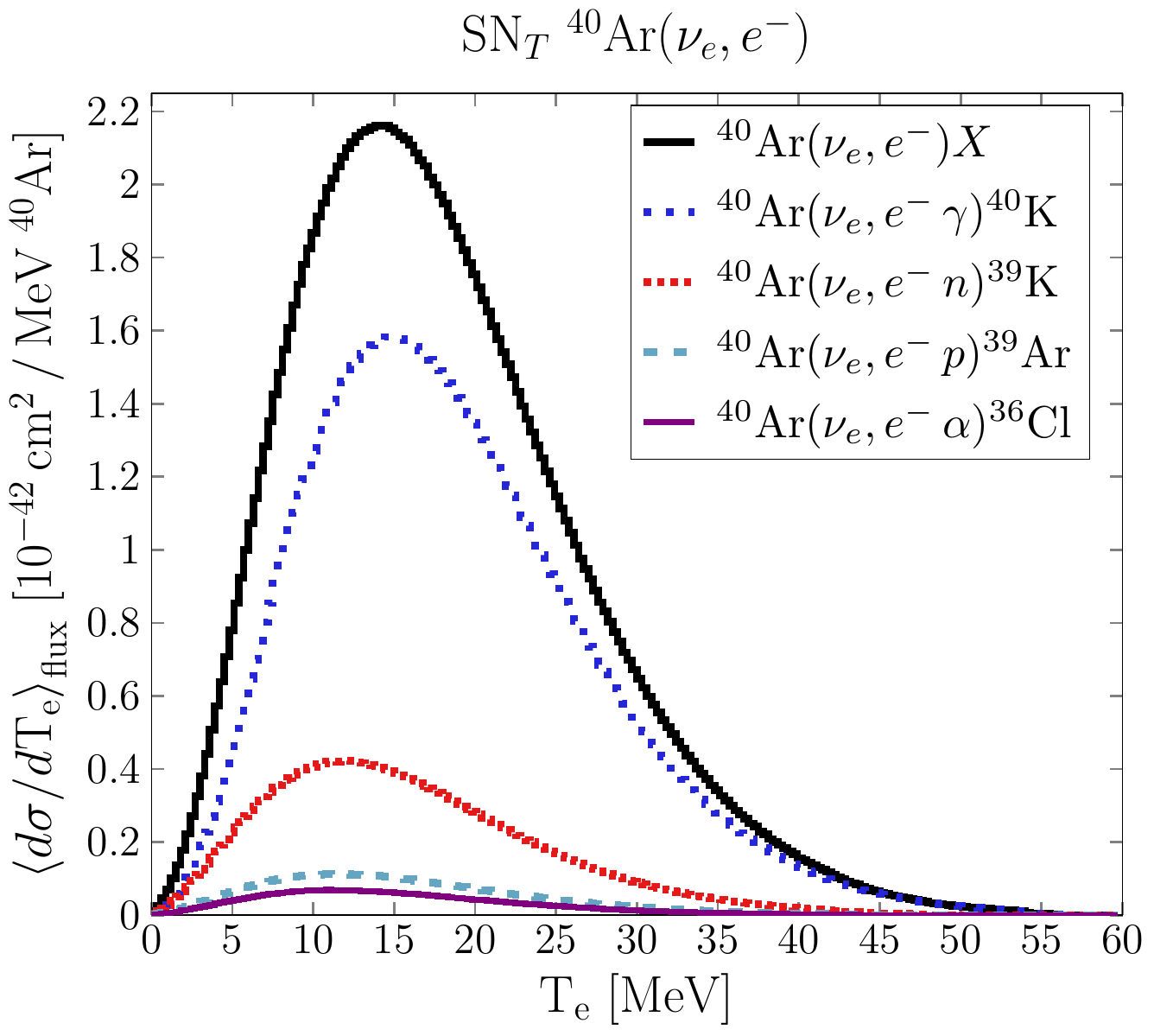}
\\[\baselineskip]
\includegraphics[width=0.99\columnwidth]{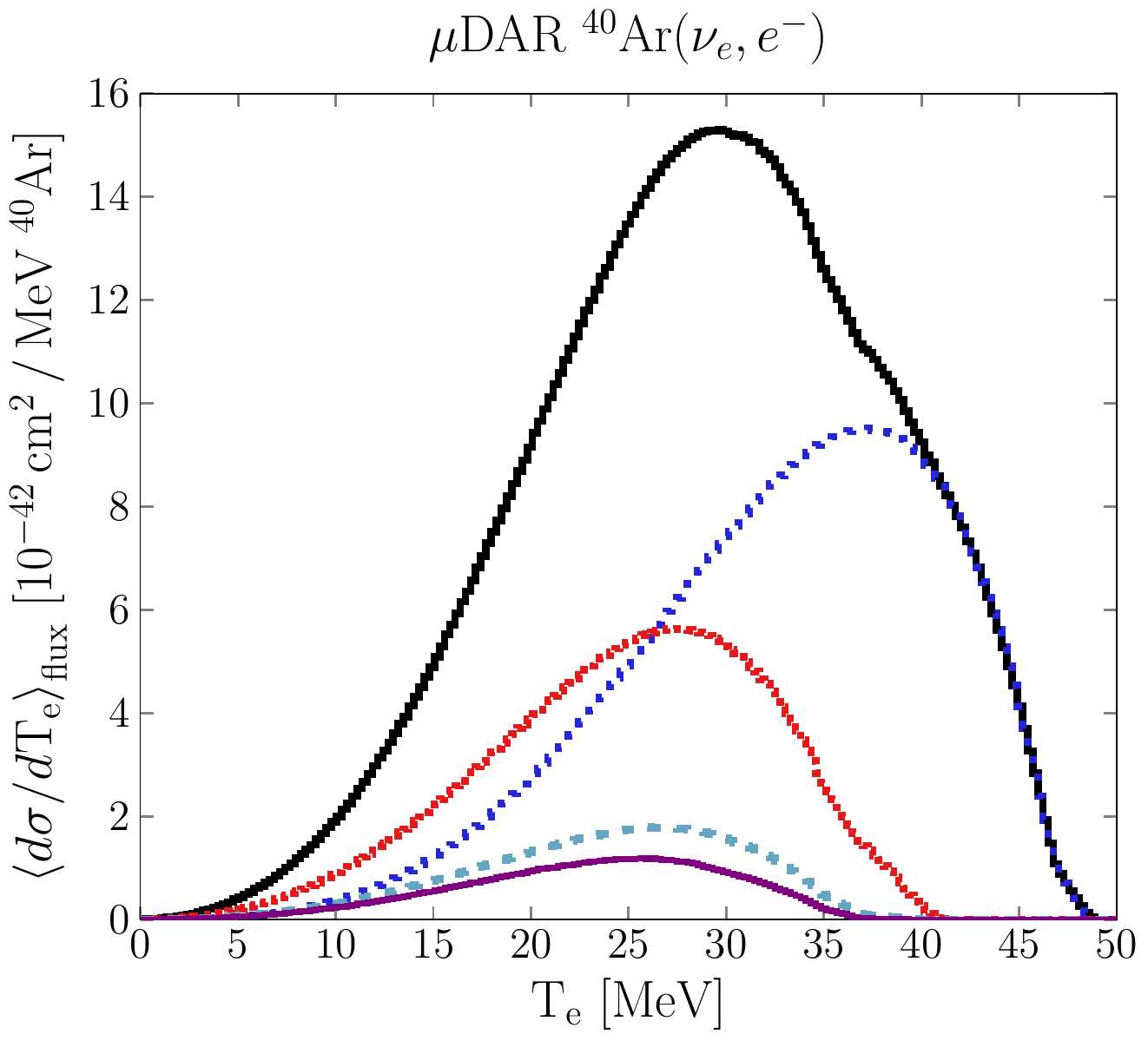}
\caption{Flux-averaged differential cross sections
with respect to the final-state electron kinetic energy
$\mathrm{T_e}$. Top: Calculation for the
time-integrated supernova $\nu_e$ spectrum described in the text.
Bottom: Calculation for $\nu_e$ produced by $\mu^{+}$ decay at rest.}
\label{fig:diff_xsecs_channels}
\end{figure}

\Cref{fig:diff_xsecs_channels} shows the flux-averaged differential cross
section with respect to the kinetic energy of the outgoing electron. The
inclusive prediction for the SN$_T$ ($\mu$DAR) spectrum is shown by the solid
black line in the upper (lower) panel, with the other line styles used to
represent individual contributions from four exclusive final states. While the
cross sections for both spectra are dominated by de-excitation modes involving
only $\gamma$-rays (loosely dotted blue) or single neutron emission (densely
dotted red), the contribution of the latter is much more pronounced for the
$\mu$DAR case. The small cross sections for single proton and single $\alpha$
emission are also noticeably enhanced as one moves from the SN$_T$ spectrum to
the $\mu$DAR spectrum.

\subsection{Neutrino energy reconstruction}
\label{sec:Ev_reco}

The energy of the incident neutrino is distributed among the
final products of the $\isotope[40]{Ar}(\nu_e,e^{-})\isotope[40]{K}^{*}$
reaction according to the relation
\begin{equation}
\label{eq:reco1}
E_\nu = E_\text{bind} + E_\mathrm{e} + \mathcal{T}_\gamma
  + \mathcal{T}_\text{ch} + \mathcal{T}_\mathrm{n} \,,
\end{equation}
where $E_\mathrm{e}$ is the total energy of the outgoing electron and
$\mathcal{T}_\gamma$, $\mathcal{T}_\text{ch}$, and $\mathcal{T}_\mathrm{n}$
are, respectively, the total kinetic energies of all de-excitation
$\gamma$-rays, charged hadrons, and neutrons in the final state.
The small recoil kinetic energy of the remnant nucleus is included
in $\mathcal{T}_\text{ch}$.
The change in binding energy $E_\text{bind}$ is given by
the expression
\begin{equation}
\label{eq:Ebind}
E_\text{bind} = \Delta_\mathrm{RT} - m_\mathrm{e}
  + \sum_\mathrm{k} m_\mathrm{k} + \sum_\mathrm{k} Q_\mathrm{k} \, m_\mathrm{e}
\end{equation}
if electron binding energies are neglected. Here
\begin{equation}
\Delta_\mathrm{RT} \equiv m_\text{atom}(\mathrm{R}) - m_\text{atom}(\mathrm{T})
\end{equation}
is the difference of ground-state atomic masses between the remnant nucleus
($\mathrm{R}$) and the nuclear target ($\mathrm{T} = \isotope[40]{Ar}$),
$m_\mathrm{e}$ is the electron mass, and $m_k$ ($Q_k$) is the mass (electric
charge) of the $k$th nuclear de-excitation product. The sums in \cref{eq:Ebind}
run over all particles emitted during nuclear de-excitations.

The minimum possible change in binding energy,
\begin{align}
E_\text{bind}^\text{min} &\equiv m_\text{atom}(\mathrm{F})
- m_\text{atom}(\mathrm{T}) - m_\mathrm{e}
\\[0.5\baselineskip]
&= m_\text{atom}(\isotope[40]{K}) - m_\text{atom}(\isotope[40]{Ar})
- m_\mathrm{e} = \SI{0.99}{\MeV}
\end{align}
occurs for final states in which only $\gamma$-rays are emitted during nuclear
de-excitations. In this case, the nuclide $\mathrm{F} = \isotope[40]{K}$
produced immediately after the primary neutrino interaction and the
nuclide $\mathrm{R}$ remaining after de-excitations are identical.

Since an a priori correction for $E_\text{bind}^\text{min}$ may be applied when
reconstructing the neutrino energy for any
$\isotope[40]{Ar}(\nu_e,e^{-})\isotope[40]{K}^{*}$ event,
\cref{eq:reco1} may be usefully rewritten in the form
\begin{equation}
\label{eq:reco2}
E_\nu = E_\text{bind}^\text{min}
  + \epsilon_\text{bind} + E_\mathrm{e} + \mathcal{T}_\gamma
  + \mathcal{T}_\text{ch} + \mathcal{T}_\mathrm{n} \,.
\end{equation}
Here I have defined the \textit{excess binding energy}
\begin{equation}
\epsilon_\text{bind} \equiv E_\text{bind} - E_\text{bind}^\text{min} \,.
\end{equation}
For \isotope[40]{K}$^*$ de-excitation modes involving only $\gamma$-rays
($\gamma$), single neutron emission ($1\mathrm{n}$), single proton emission
($1\mathrm{p}$), or the emission of both a single neutron and a single proton
($1\mathrm{n}1\mathrm{p}$), the excess binding energy takes the values
\begin{align}
\epsilon_\text{bind}^\gamma &= 0
\\[0.5\baselineskip]
\epsilon_\text{bind}^\mathrm{1n} &= \SI{7.80}{\MeV}
\\[0.5\baselineskip]
\epsilon_\text{bind}^\mathrm{1p} &= \SI{7.58}{\MeV}
\\[0.5\baselineskip]
\epsilon_\text{bind}^\mathrm{1n1p} &= \SI{14.18}{\MeV} \,.
\end{align}

A useful property of the excess binding energy is that only a few discrete
values of this variable are likely to occur at supernova energies. A future
analysis of supernova neutrino data may therefore attempt to correct for
nonzero values of $\epsilon_\text{bind}$ by tagging events in which a nucleon
or a heavy fragment was emitted from the struck nucleus.

Beyond the binding energy contributions, the other terms in \cref{eq:reco2}
vary in the degree to which they may be reconstructed by a detector. In a
liquid argon time projection chamber (LArTPC), the primary electron will
produce a \si{\centi\meter}-scale ionization track which may be used to
determine its energy and direction. De-excitation $\gamma$-rays will produce
isolated small energy depositions within several tens of \si{\centi\meter} of
the interaction vertex, primarily via Compton scattering on atomic electrons.
Reconstruction of both of these features for supernova neutrino interactions is
considered in ref.~\cite{Castiglioni2020}, with the conclusion that the energy
associated with each can largely be recovered under realistic detector
performance assumptions. Neutron tagging and calorimetry, on the other
hand, were found to be far more challenging.

Low-energy charged nuclear fragments, such as protons and alpha particles, may
also produce observable ionizations in a LArTPC. A key challenge for
identifying the activity induced by these particles is that, at the energies
relevant for supernova neutrinos, charged hadrons will produce
\si{\milli\meter}-scale or smaller ionization tracks. These will likely be
difficult to distinguish from the longer track produced by the primary
electron. However, if events involving charged nuclear fragment emission can be
successfully tagged, perhaps by identifying unusually large charge deposits
near the start of the primary electron track, then at least some of the charged
hadron kinetic energy may be recoverable.

To assess the relative importance of the various terms on the right-hand side
of \cref{eq:reco2}, I define several observables, all of which may be
interpreted as a reconstructed neutrino energy $E_\nu^\text{reco}$ under
different, often quite optimistic, assumptions. The simplest reconstruction
method involves adding the outgoing electron's total energy to the
minimum possible change in binding energy for CC $\nu_e$ absorption on
\isotope[40]{Ar}:
\begin{equation}
E_\mathrm{e}^\text{reco} \equiv E_\text{bind}^\text{min} + E_\mathrm{e}\,.
\end{equation}
This estimate of the neutrino energy may be refined by adding the summed
energies of all de-excitation $\gamma$-rays
\begin{equation}
E_{\mathrm{e}+\gamma}^\text{reco} \equiv E_\mathrm{e}^\text{reco}
  + \mathcal{T}_\gamma
\end{equation}
and further refined by adding the summed kinetic energies of all
final-state charged hadrons
\begin{equation}
E_\text{vis}^\text{reco} \equiv E_{\mathrm{e}+\gamma}^\text{reco}
  + \mathcal{T}_\mathrm{ch} \,.
\end{equation}
I call the last of these variables the \textit{visible energy} while
recognizing that low-energy neutrons may nevertheless produce some observable
signals in a LArTPC.

Finally, I consider three possible strategies for implementing a binding energy
correction via tagging of final-state nuclear fragments. All three involve
conditionally adding one or more terms to the expression for the visible energy
above. Under the assumption that de-excitation neutrons may be successfully
tagged, I define the reconstructed neutrino energy
\begin{equation}
E_\text{tag}^\text{n} \equiv E_\text{vis}^\text{reco}
  + \delta_\mathrm{n} \, \epsilon_\text{bind}^{1\mathrm{n}}
\end{equation}
in which the symbol $\delta_\mathrm{n}$ is defined to be unity when
a \marley\ event contains at least one final-state neutron and
zero otherwise. Similarly, under the assumption that charged
nuclear fragment emission may be successfully identified,
I define
\begin{equation}
E_\text{tag}^\text{p} \equiv E_\text{vis}^\text{reco}
  + \delta_\mathrm{ch} \, \epsilon_\text{bind}^{1\mathrm{p}}
\end{equation}
in which $\delta_\mathrm{ch}$ is unity when a \marley\ event contains a charged
nuclear fragment in the final state and zero when it does not. In an ideal
scenario in which both of these tagging techniques are reliable, a still more
refined estimate of the neutrino energy may be obtained via
\begin{align}
\nonumber
E_\text{tag}^{\text{n}+\text{p}} \equiv E_\text{vis}^\text{reco}
  &+ \delta_\mathrm{n} \, (1 - \delta_\mathrm{ch}) \,
    \epsilon_\text{bind}^{1\mathrm{n}}
\\[0.5\baselineskip]
  &+ \delta_\mathrm{ch} \, (1 - \delta_\mathrm{n}) \,
    \epsilon_\text{bind}^{1\mathrm{p}}
  + \delta_\mathrm{n} \, \delta_\mathrm{ch} \,
    \epsilon_\text{bind}^{1\mathrm{n}1\mathrm{p}} \,.
\end{align}

\begin{figure}
\includegraphics[width=0.99\columnwidth]{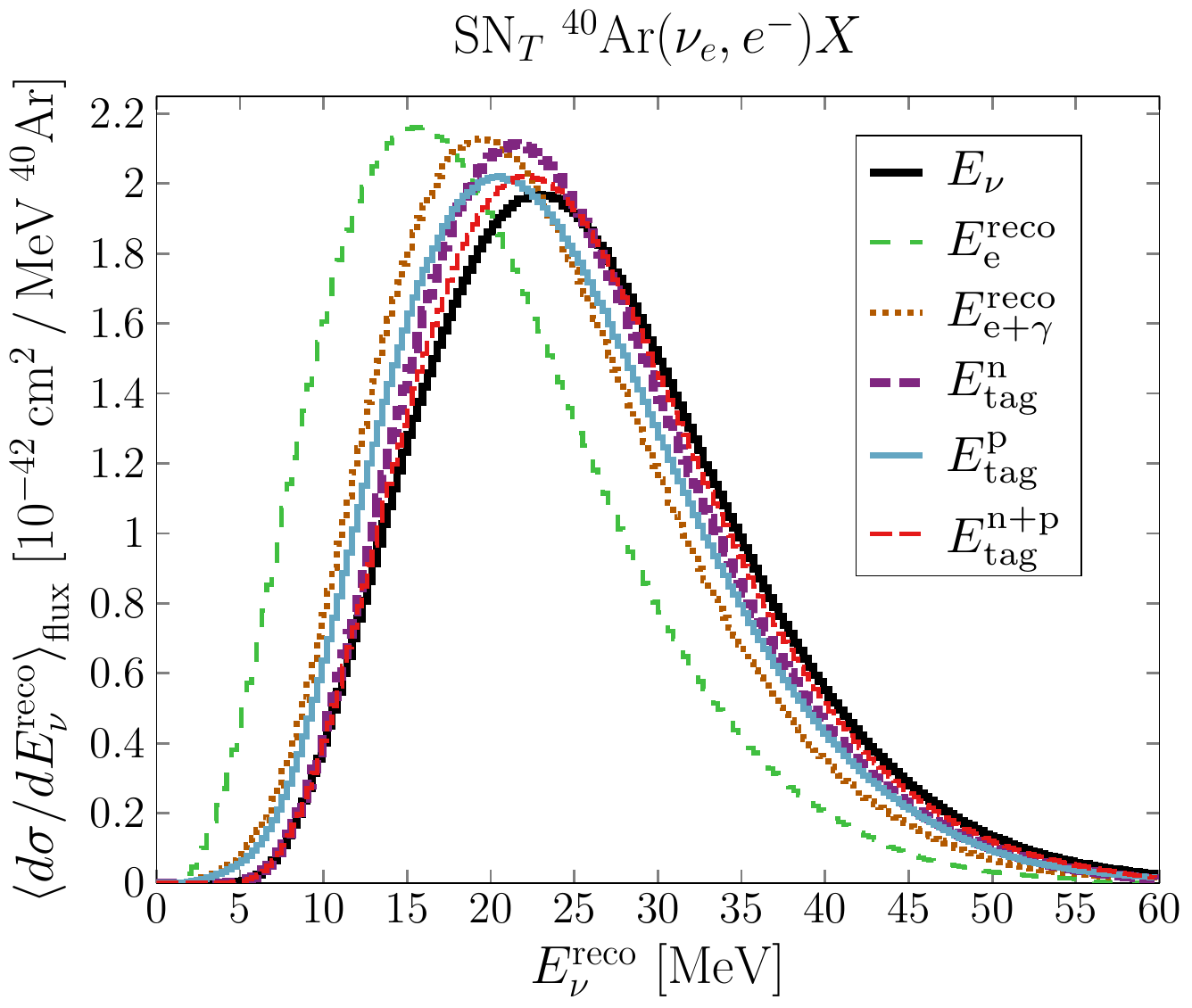}
\\[0.5\baselineskip]
\includegraphics[width=0.99\columnwidth]{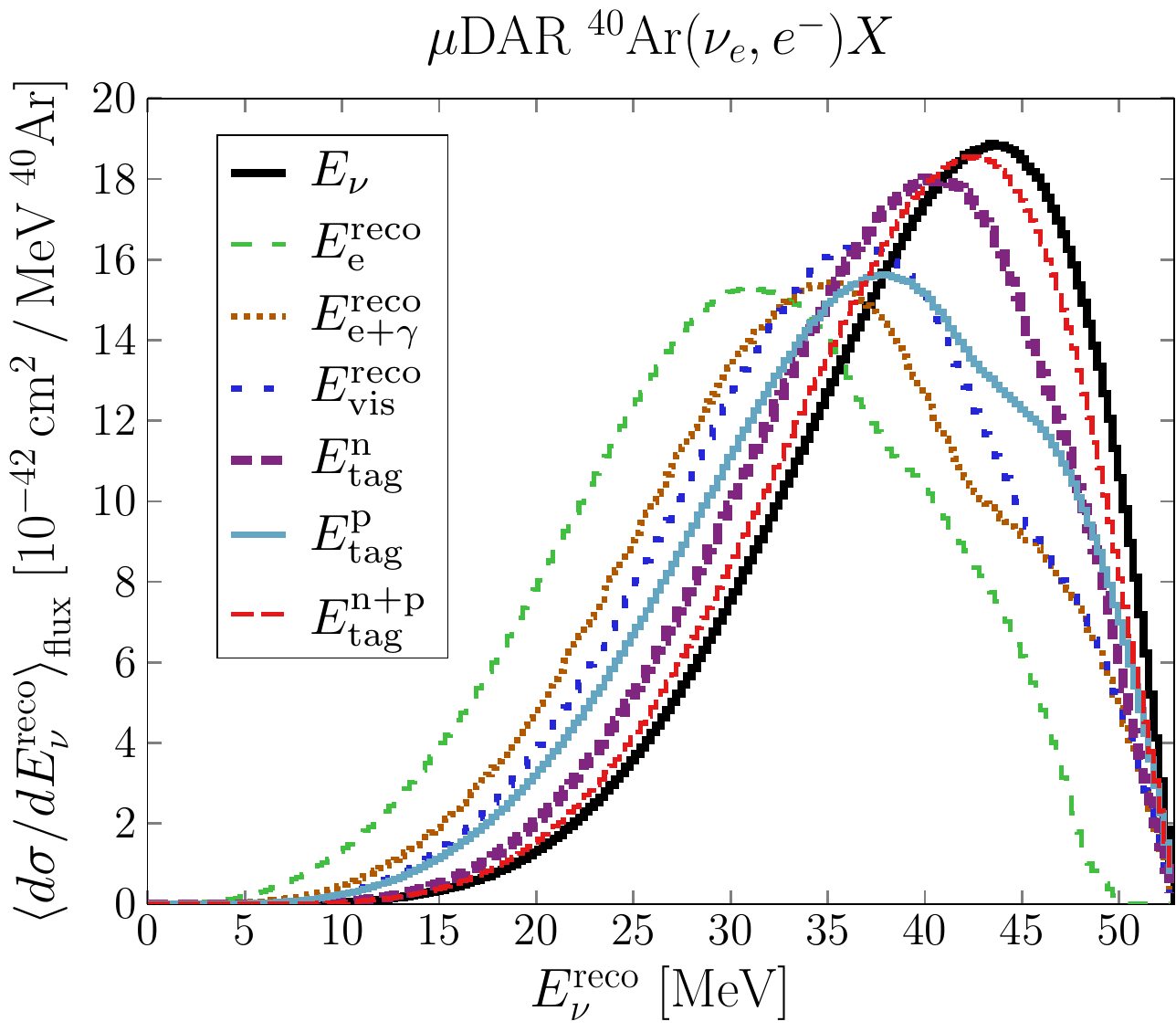}
\caption{Flux-averaged differential cross sections
with respect to various definitions of the reconstructed
neutrino energy $E_\nu^\text{reco}$.
Top: Calculation for the
time-integrated supernova $\nu_e$ spectrum described in the text.
Bottom: Calculation for $\nu_e$ produced by $\mu^{+}$ decay at rest.}
\label{fig:diff_xsecs_2}
\end{figure}

\Cref{fig:diff_xsecs_2} shows the \marley\ prediction for flux-averaged
differential sections with respect to each of these observables. The top
(bottom) panel of the figure presents results for the SN$_T$ ($\mu$DAR) energy
spectrum defined earlier. A solid black line is used to draw the differential
cross section with respect to the true neutrino energy, while the other line
styles represent the various methods for reconstructing it. The
$E_\text{vis}^\text{reco}$ result is not shown in the top panel since it is
difficult to distinguish from the $E_{\mathrm{e}+\gamma}^\text{reco}$ one on
the scale of the plot.

For both spectra studied, the agreement between the reconstructed and true
neutrino energy distributions improves most dramatically as one moves from
using only the primary electron ($E_\mathrm{e}^\text{reco}$, thin dashed green)
to using both the electron and the de-excitation $\gamma$-rays
($E_{\mathrm{e}+\gamma}^\text{reco}$, densely dotted brown) in the
reconstruction. Although inclusion of information about charged hadrons is also
seen to be helpful, the next most important improvement comes from the
inclusion of binding energy corrections related to neutron tagging
($E_\text{tag}^\mathrm{n}$, thick dashed violet). Due to the higher mean energy
of the $\mu$DAR spectrum, nuclear fragment emission becomes more important
relative to SN$_T$, and the impact of the tagging-based binding energy
corrections on neutrino energy reconstruction becomes more pronounced.

\begin{figure}
\includegraphics[width=0.99\columnwidth]{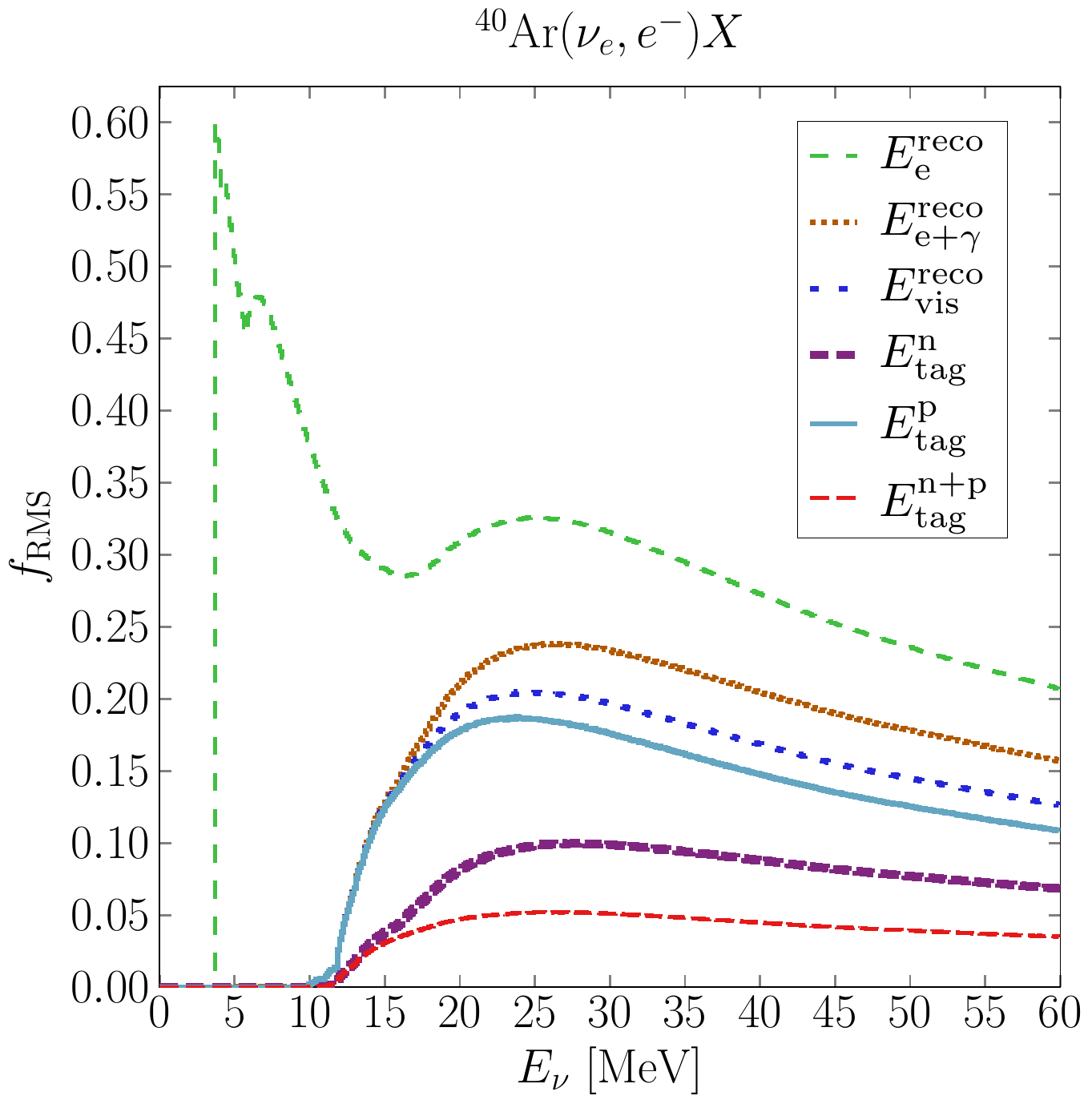}
\caption{Fractional RMS resolution for the neutrino energy reconstruction
methods described in the text.}
\label{fig:frac_rms_resolution}
\end{figure}

To further quantify the performance of each of these energy reconstruction
methods, \cref{fig:frac_rms_resolution} reports the fractional root mean
square (RMS) resolution
\begin{equation}
\label{eq:frac_rms}
f_\text{RMS}(E_\nu) \equiv
\sqrt{ \left< \left(\frac{E_\nu^\text{reco} - E_\nu }
{ E_\nu }\right)^{\!2} \right> }
\end{equation}
for each definition of the reconstructed neutrino energy $E_\nu^\text{reco}$
 above. Here, the angle brackets denote the arithmetic mean of the
enclosed quantity. The choice of $f_\text{RMS}$ as a metric is intended to
facilitate comparisons with fig.~4 of ref.~\cite{Castiglioni2020} and fig.~7 of
ref.~\cite{Abi2020}, both of which use the same quantity to study energy
reconstruction in a full LArTPC detector simulation. The results are shown in
small bins of the true neutrino energy $E_\nu$. To obtain the curves shown
in \cref{fig:frac_rms_resolution}, a large sample of \marley\ events was
generated, and a Monte Carlo estimator for
$f_\text{RMS}$ was evaluated via
\begin{equation}
f_\text{RMS}(E_\nu \in b) \approx
\sqrt{ \frac{1}{N_b} \, \sum_{j=1}^{N_b}
\left[\frac{ E_\nu^\text{reco}(j) - E_\nu(j) }{ E_\nu(j) }\right]^2 }
\end{equation}
where the sum runs over over the $N_b$ simulated events which fell into
the neutrino energy bin $b$ of interest.

The vertical line seen around \SI{4}{\MeV} for the $E_\mathrm{e}^\text{reco}$
curve (thin dashed green) corresponds to the \marley\ energy threshold for CC
$\nu_e$ absorption. Because the third-forbidden transition between the ground
states of \isotope[40]{Ar} ($J^\pi = 0^{+}$) and \isotope[40]{K} ($J^\pi =
4^{-}$) is neglected under the allowed approximation, \marley\ predicts a
finite resolution for $E_\mathrm{e}^\text{reco}$ even at threshold.

The considerable improvements in energy resolution seen between
$E_\mathrm{e}^\text{reco}$ and $E_{\mathrm{e}+\gamma}^\text{reco}$ (densely
dotted brown) and between $E_{\mathrm{e}+\gamma}^\text{reco}$ and
$E_\text{tag}^\mathrm{n}$ (thick dashed violet) further highlight the
conclusions mentioned above with respect to \cref{fig:diff_xsecs_2}: while
$\nu_e$ energy reconstruction in CC absorption on argon will benefit from
increased information about any final-state particle species, determining the
de-excitation $\gamma$-ray energies and tagging neutrons are both particularly
impactful.

\section{Summary and conclusions}
\label{sec:conclusions}

Due to the potential for DUNE to obtain a once-in-a-lifetime large-statistics
measurement of supernova electron neutrinos, achieving a detailed understanding
of tens-of-\si{\MeV} neutrino-argon interactions is an investment that may
yield a high scientific return. This paper expands the ability to model these
interactions by providing a first calculation of exclusive cross sections for
the $\isotope[40]{Ar}(\nu_e,e^{-})\isotope[40]{K}^{*}$ reaction at supernova
energies. The implementation of the models underlying this calculation in the
\marley\ event generator enables studies of neutrino energy reconstruction to
be carried out easily. The simple approach pursued in \cref{sec:Ev_reco}
reveals the substantial role that measuring the energies of de-excitation
$\gamma$-rays and (though difficult) neutron tagging may play in optimizing
supernova $\nu_e$ energy resolution in a future analysis by DUNE. Further
insights are available by using \marley\ in conjunction with a realistic
detector simulation \cite{Abi2020,Castiglioni2020}.

Two major approximations adopted in \marley\ \version\ constitute limitations
on the present study that should be revisited in future research. The first of
these is the allowed approximation invoked during derivation of the inclusive
differential cross section in \cref{sec:allowed_approx}. In a more detailed
calculation of this cross section, the factor
$e^{i\mathbf{q}\cdot\mathbf{x}(n)}$ that appears in the nuclear matrix element
from \cref{eq:nuclear_matrix_element} is expanded in a series of multipoles
\cite{Walecka1975} that depend on the spherical Bessel function
$j_J(|\mathbf{q}|\,r_n)$, where $r_n$ is the magnitude of $\mathbf{x}(n)$ and
$J$ is the multipole order. Terms representing forbidden nuclear transitions
($J > 0$) vanish in the $|\mathbf{q}| \to 0$ limit imposed by the allowed
approximation, but their contribution to the cross section becomes increasingly
important as the momentum transfer grows. Since the centroid energy of the
multipole giant resonances grows with $J$ roughly like $41\,J\,A^{-1/3}$
\si{\MeV} \cite{Kolbe2003a}, the inclusion of forbidden transitions should
enhance neutrino scattering to high-lying unbound nuclear states which
de-excite primarily via fragment emission. The degree to which this observation
affects the present results may be studied in the future by combining a more
detailed calculation of the inclusive differential cross section with the
\marley\ de-excitation model.

The second major approximation used in this work, which is shared by nearly all
calculations of exclusive cross sections for tens-of-\si{\MeV} neutrino-nucleus
scattering, is the compound nucleus assumption discussed in \cref{sec:unbound}.
Further investigation, both theoretical and experimental, will be needed to
clarify the degree to which direct knock-out and pre-equilibrium processes may
safely be neglected in models of low-energy neutrino-nucleus reactions. A key
question is how the transition between the compound nucleus picture, which is
standard for low-energy neutrinos, and the intranuclear cascade picture, which
is commonly used in models of accelerator neutrino interactions, should be
handled as a function of neutrino energy.

Although the current discussion has focused specifically on the description of
nuclear de-excitations following CC $\nu_e$ absorption on \isotope[40]{Ar}, the
model presented in \cref{sec:deex_model} is sufficiently general that it may be
applied unaltered in a variety of other contexts. A natural next step is the
use of \marley\ together with an inclusive description of inelastic
neutral-current scattering on argon, a process for which de-excitations provide
the only experimental observables apart from nuclear recoil. While measurements
of tens-of-\si{\MeV} neutrino-argon inelastic cross sections must be pursued to
meet the needs of the DUNE supernova neutrino program, more immediate
opportunities for confronting \marley\ with data may become available if the
code is used to obtain predictions for other nuclei. Near-future measurements
that could provide a detailed test of \marley\ include studies of CC $\nu_e$
absorption on carbon by JSNS$^2$ \cite{Ajimura2017} and neutrino-induced
neutron production on lead, iron, and copper by COHERENT \cite{COH2018}.
Measurements of exclusive cross sections and decay rates for processes that are
closely related to neutrino interactions, such as electron-nucleus scattering
\cite{Flowers1978} and muon capture \cite{MuonArgon1,Parvu2018}, may also
provide helpful model constraints. Finally, the capabilities of \marley\ may
prove useful in simulating nuclear de-excitations induced by processes beyond
the Standard Model, including nucleon decay \cite{NDK1,NDK2} and the absorption
of fermionic dark matter \cite{DM1,DM2}.

\section{Acknowledgements}

I thank Myung-Ki Cheoun for providing the QRPA matrix elements used in this
work, and I am grateful to the TALYS authors for the decision to release their
nuclear structure data files under the terms of the GNU General Public License.

Robert Svoboda, Ramona Vogt, and Michael Mulhearn provided helpful feedback on
the doctoral thesis \cite{Gardiner2018} that was a precursor to this
publication.

I gratefully acknowledge financial support while at the University of
California, Davis from the John Jungerman-Charles Soderquist Graduate
Fellowship and from the DOE National Nuclear Security Administration through
the Nuclear Science and Security Consortium under award number DE-NA0003180.

This manuscript has been authored by Fermi Research Alliance, LLC under
Contract No. DE-AC02-07CH11359 with the U.S. Department of Energy, Office of
Science, Office of High Energy Physics.

The high-statistics \marley\ calculations reported herein were performed using
resources provided by the Open Science Grid, which is supported by the National
Science Foundation and the U.S. Department of Energy's Office of Science.

\appendix

\section{Decay width for fragment emission}
\label{sec:derive_width}

In this appendix, I give a brief derivation of the expression in
\cref{eq:fragment_diff_decay_width} for the nuclear fragment emission
differential decay width of a compound nucleus. A similar approach can be used
to obtain the result in \cref{eq:gamma_diff_decay_width} for $\gamma$-ray
emission. The argument presented here is a modern version of one originally
given by Weisskopf in ref.~\cite{Weisskopf1937}.

Consider the decay process $i \to a + f$ in which a compound nucleus $i$ emits
a fragment $a$ to become a final-state nucleus $f$. Adopt the same notation as
in \cref{sec:diff_decay_widths}: the initial (final) nucleus has spin $J$
($J^\prime$), parity $\Pi$ ($\Pi^\prime$), and mass $M$ ($M^\prime$). The
emitted fragment has spin $s$, orbital (total) angular momentum $\ell$ ($j$),
three-momentum magnitude $\fragmentMomCM$, mass $m_a$, and parity $\pi_a$.
Denote the initial nuclear excitation energy by $E_x$, and let the final
nuclear excitation energy lie on the small interval $[ E_x^\prime, E_x^\prime +
dE_x^\prime ]$. The symbol $\rho_i$ ($\rho_f$) denotes the spin-parity
dependent level density (see \cref{sec:density_spin}) in the vicinity of $E_x$
($E_x^\prime$) for the initial (final) nucleus.

Within an arbitrary volume $V$ and in the rest frame of the initial nucleus,
the number of states $n_{a+f}$ that may be populated by the decay is given by
\begin{equation}
n_{a+f} = n_a \, n_f
\end{equation}
where
\begin{equation}
n_a = (2s + 1) \, \frac{V}{2\pi^2}
\, \fragmentMomCM^2 \left| \frac{ d\fragmentMomCM }
{ dE_x^\prime } \right| dE_x^\prime
\end{equation}
and
\begin{equation}
n_f = (2J^\prime + 1) \, \rho_f(E_x^\prime, J^\prime, \Pi^\prime)
\, dE_x^\prime \,.
\end{equation}

By detailed balance, the decay width $\Gamma_{a+f}$
is related to the width $\Gamma_i$ of the time-reversed absorption
process $a + f \to i$ via
\begin{equation}
\Gamma_{a+f} = \frac{ n_{a+f} }{ n_i } \, \Gamma_i
\end{equation}
where
\begin{equation}
n_i = (2J + 1) \, \rho_i(E_x, J, \Pi) \, \frac{ dE_x }
{ dE_x^\prime } \, dE_x^\prime
\end{equation}
is the number of states in which the compound nucleus $i$ may be formed.
The absorption width may be written as
\begin{equation}
\Gamma_i = \phi \, \sigma
\end{equation}
where
\begin{equation}
\phi = \frac{ M \, \fragmentMomCM } { V \, E_a \, E_f }
= \frac{ 1 }{ V } \frac{ dE_x }{ d\fragmentMomCM }
\end{equation}
is the particle flux and
\begin{equation}
\label{eq:cnfxsec}
\sigma = \frac{ \pi \, (2J + 1) }{ \fragmentMomCM^2 \, (2s + 1) (2J^\prime + 1) }
\sum_{\ell = 0}^\infty \; \sum_{j = |\ell - s|}^{\ell + s}
T_{\ell j}(\epsilon)
\end{equation}
is the compound nucleus formation cross section.
Here $E_a$ ($E_f$) is the total energy of the emitted fragment
(final nucleus) and
\begin{equation}
\epsilon = M - m_a - M^\prime
\end{equation}
is the total kinetic energy of the $a + f$ system. A derivation of the
expression in \cref{eq:cnfxsec} is given\footnote{Note that there is a
misprint in equations (A.5) and (A.6) from ref.~\cite{Gardiner2018}. The
quantities $\ell \pm s_A$, $\ell^\prime \pm s_B$, and $\ell \pm s_b$
should be replaced wherever they occur with, respectively, $j \pm s_A$, $j^\prime \pm s_B$, and $\ell^\prime
\pm s_b$.} in ref.~\cite{Gardiner2018}.
Similar derivations can also be found in, e.g.,
refs.~\cite{Cole2000,Thompson2009}.

Combining the results above and summing over the allowed values of
$J^\prime$, which satisfies the triangle relation
\begin{equation}
|J - j| \leq J^\prime \leq J + j \,,
\end{equation}
leads immediately to \cref{eq:fragment_diff_decay_width}.

\section{Level density model}
\label{sec:level_density_model}

The nuclear level density model used in the present calculations is the
RIPL-3 parameterization \cite{RIPL3} of the Back-shifted Fermi gas Model
(BFM), which is based on the work of Koning, Hilaire, and Goriely
\cite{Koning2008}. The ``back shift'' used by this model, which accounts for
nucleon pairing effects, involves replacing the nuclear excitation energy $E_x$
by an effective value $U$ defined by
\begin{equation}
U \equiv E_x - \Delta
\end{equation}
where the energy shift
\begin{equation}
\label{eq:BFM_energy_shift}
\Delta = \chi_\text{pair}\, \frac{ \SI{12}{\MeV} }{ \sqrt{A} } + \delta
\end{equation}
is adjusted to fit experimental data using the empirical parameter $\delta$.
The pairing factor $\chi_\text{pair}$ is defined by
\begin{equation}
\chi_\text{pair} \equiv \begin{cases} 1 & \text{even-even}
\\ 0 & \text{odd-$A$}
\\ -1 & \text{odd-odd}.
\end{cases}
\end{equation}

\subsection{Total level density}

Under the Back-shifted Fermi gas model (BFM), the total density of nuclear
levels near excitation energy $E_x$ is given by the expression
\cite{Koning2008}
\begin{equation}
\rho(E_x) = \left[ \frac{1}{ \rho_F(E_x) }
+ \frac{1}{ \rho_0(E_x) }  \right]^{-1}
\end{equation}
where
\begin{equation}
\rho_F(E_x) \equiv \frac{1}{\sqrt{2\pi} \sigma} \, \frac{\sqrt{\pi}}{12}
  \, \frac{ \exp(2\sqrt{a_{\rm LD} \, U}) }{ a_{\rm LD}^{1/4} \, U^{5/4} }
\end{equation}
is the Fermi gas level density and
\begin{equation}
\rho_0(E_x) = \frac{a_{
\rm LD}}{12\,\sigma} \, \exp(a_{\rm LD} \, U + 1)
\end{equation}
is a correction intended to suppress the unphysical divergence of
$\rho_F(E_x)$ at low excitation energies.

Although a constant value for the level density parameter $a_{\rm LD}$ is
sometimes used, I adopt the energy-dependent functional form
\cite{Ignatyuk1975} recommended by RIPL-3 to correct for the damping of shell
effects at high excitation energies:
\begin{align}
\nonumber
a_{
\rm LD} &\equiv a_{
\rm LD}(E_x, Z, A)
\\[0.5\baselineskip]
& = \tilde{a}(A)\left\{ 1 + \frac{\delta W(Z,A)}{U}
\left[ 1 - \exp(-\gamma \, U) \right] \right\}.
\end{align}
Here $\delta W(Z,A)$ is the shell correction energy, $\tilde{a}(A)$
is the asymptotic value of $a_{
\rm LD}$ at high excitation energies, and $\gamma$
is a damping parameter that represents how quickly $a_{
\rm LD}(E_x, Z, A)$ approaches
$\tilde{a}(A)$. The values of these three parameters are given by the
relations
\begin{align}
\delta W(Z,A) &= \delta M_\text{exp}(Z,A) - \delta M_\text{LDM}(Z, A)
\\[0.5\baselineskip]
\tilde{a} &= \alpha \, A + \beta \, A^{2/3}
\\[0.5\baselineskip]
\gamma &= \gamma_0 \, A^{-1/3}
\end{align}
where $\delta M_\text{exp}(Z,A)$ is the measured nuclear mass excess
\cite{AME2012, AME2012b} for the nuclide with proton number $Z$ and mass
number $A$, and $\delta M_\text{LDM}(Z,A)$ is the corresponding prediction for
the nuclear mass excess using the liquid drop model
\cite[p. 3164]{RIPL3} \cite{Mengoni1994,Myers1966}.

This work uses the global ``BFM effective'' values of the empirical parameters
$\alpha$, $\beta$, $\delta$, and $\gamma_0$ obtained in ref.~\cite{Koning2008}
using fits to nuclear level data:
\begin{align}
\alpha &= \SI{0.0722396}{\per\MeV} & \beta &= \SI{0.195267}{\per\MeV}
\\[0.5\baselineskip]
\gamma_0 &= \SI{0.410289}{\per\MeV} & \delta &= \SI{0.173015}{\MeV}.
\end{align}

\subsection{Spin dependence}
\label{sec:density_spin}

The density of nuclear levels $\rho(E_x, J, \Pi)$ with total spin $J$ and
parity $\Pi$ near excitation energy $E_x$ may be written in the form
\begin{equation}
\rho(E_x, J, \Pi) = \pi(E_x, J, \Pi) \, R(E_x, J) \, \rho(E_x)
\end{equation}
where $R(E_x, J)$ is the nuclear spin distribution and $\pi(E_x, J, \Pi)$
is the parity distribution.

Under the assumption that the individual nucleon spins are pointing in random
directions, it can be shown \cite{Bethe1937} that the spin distribution
$R(E_x,J)$ is given by \cite{Koning2008}
\begin{equation}
R(E_x, J) = \frac{2J+1}{2\,\sigma^2} \exp\left[ - \frac{ (J+\frac{1}{2})^2 }
{2\,\sigma^2} \right].
\end{equation}
The spin cutoff parameter $\sigma$ determines the width of $R(E_x,J)$.
To calculate this parameter, I adopt the expression recommended
by RIPL-3 \cite{RIPL3} in the absence of discrete level data:
\begin{align}
\nonumber
\sigma^2 &= \sigma^2(E_x)
\\[0.5\baselineskip]
&= \begin{cases}
\sigma_{d,\text{global}}^2 + \frac{E_x}{S_n}\left[ \sigma^2_F(S_n) -
\sigma_{d,\text{global}}^2 \right] & \text{for } E_x < S_n \\
\sigma^2_F(E_x) & \text{for } E_x \geq S_n.
\end{cases}
\end{align}
Here $S_n$ is the neutron separation energy,
\begin{equation}
\sigma^2_F(E_x) \equiv \left( \SI{0.01389}{\per\MeV} \right)
\frac{A^{5/3}}{\tilde{a}} \sqrt{a_{\rm LD} \, U} \,,
\end{equation}
and
\begin{equation}
\sigma_{d,\text{global}} \equiv 0.83\,A^{0.26} \,.
\end{equation}

\subsection{Parity dependence}

Most level density calculations assume equipartition of parity, i.e.,
\begin{equation}
\pi(E_x, J, \Pi) = \frac{1}{2} \,.
\end{equation}
I adopt this assumption in agreement with RIPL-3. However, I note that more
sophisticated treatments of $\pi(E_x, J, \Pi)$ have been proposed (see, e.g.,
ref.~\cite{AlQuraishi2003}).

\section{Optical potential}
\label{sec:optical_potential}

For the statistical model calculations reported here and implemented in
\marley, the global nuclear optical potential developed by Koning and Delaroche
\cite{Koning2003} has been adopted. This phenomenological potential is based on
fits to nucleon-nucleus scattering data and may be written in the form
\begin{align}
\mathcal{U} =& -\mathcal{V}_V - i \, \mathcal{W}_V
- i \, \mathcal{W}_D + d_{\ell s} \,
\big(\mathcal{V}_{SO} + i \, \mathcal{W}_{SO}\big)
+ \mathcal{V}_C \,.
\end{align}
Here
\begin{equation}
d_{\ell s} \equiv j(j+1) - \ell(\ell+1) - s(s+1).
\end{equation}
is the eigenvalue of the spin-orbit operator $2 \, \pmb{\ell} \cdot \pmb{s}$
for a projectile with definite total angular momentum $j$, orbital angular
momentum $\ell$, and spin $s$.

The Coulomb potential $\mathcal{V}_C$ is taken to be that of a
uniformly-charged sphere:
\begin{equation}
\mathcal{V}_C(r) =
  \begin{cases}
    \frac{Z\,z\,e^2}{2 \, R_C}\left( 3 - \frac{r^2}{R_C^2} \right) & r < R_C
    \\[0.3cm]
    \frac{Z\,z\,e^2}{r} & r \geq R_C
  \end{cases}
\end{equation}
In the expression above, $r$ is the radial coordinate of the projectile, $Z$
($z$) is the proton number of the target nucleus (projectile), $e$ is the
elementary charge, and $R_C$ is the Coulomb radius of the nucleus.

\subsection{Nucleon projectiles}

The volume ($V$), surface ($D$), and spin-orbit ($SO$) terms of the optical
potential are functions that may be expressed as the product of an
energy-dependent well depth and an energy-independent radial part:
\begin{align}
\label{eq:start_KD_well_depths}
\mathcal{V}_V &= V_V(\fragmentKinELab) \, f(r,R_V,a_V)
\\[0.5\baselineskip]
\mathcal{W}_V &= W_V(\fragmentKinELab) \, f(r,R_V,a_V)
\\[0.5\baselineskip]
\mathcal{W}_D &= -4 \, a_D^\NucleonSpecies \, W_D(\fragmentKinELab)
  \, \frac{d}{dr}f(r,R_D,a_D^\NucleonSpecies)
\\[0.5\baselineskip]
\mathcal{V}_{SO} &= V_{SO}(\fragmentKinELab) \, \frac{1}{m_\pi^2 \, r}
\, \frac{d}{dr}f(r,R_{SO},a_{SO})
\\[0.5\baselineskip]
\label{eq:end_KD_well_depths}
\mathcal{W}_{SO} &= W_{SO}(\fragmentKinELab) \, \frac{1}{ m_\pi^2 \, r }
\, \frac{d}{dr} f(r,R_{SO},a_{SO})
\end{align}
Here $m_\pi$ is the mass of a charged pion, and the well depths $V_V$, $W_V$,
etc., are real-valued functions of the laboratory kinetic energy
$\fragmentKinELab$ of the projectile.

The radial dependence in \crefrange{eq:start_KD_well_depths}
{eq:end_KD_well_depths} is given by the Woods-Saxon \cite{Woods1954} shape
\begin{equation}
f(r, R, a) = \left(1 + \exp\left[ (r-R)/a \right]\right)^{-1}
\end{equation}
with effective radius $R$ and diffuseness parameter $a$.
\mbox{\Cref{tab:optical_model_params_radial}} lists the values of the parameters
needed to compute the radially-dependent parts of the nuclear optical
potential. Each effective radius $R_j$ is related to its tabulated parameter
$r_j$ via
\begin{align}
R_j &= r_j \, A^{1/3} & j \in \{V,D,SO,C\}\,.
\end{align}
Note that all parameter values listed in \cref{tab:optical_model_params_radial}
are given in \si{\femto\meter}, while the expressions given in the text assume
natural units ($\hbar = c = 1$).

The expressions for the well depths are most conveniently written in terms of
\begin{equation}
\FermiEDiff \equiv \fragmentKinELab - E_F^\NucleonSpecies \,,
\end{equation}
the difference between the laboratory-frame kinetic energy of the projectile
$\fragmentKinELab$ and the nuclear Fermi energy $E_F^\NucleonSpecies$
for the projectile species $\NucleonSpecies \in
\mathrm{\{p,n\}}$ of interest:
\begin{align}
V_V(\fragmentKinELab) &= v_1^\NucleonSpecies \big( 1
- v_2^\NucleonSpecies \, \FermiEDiff
+ v_3^\NucleonSpecies \, \FermiEDiff^2
- v_4^\NucleonSpecies \, \FermiEDiff^3 \big) + V_\text{Coul}
\\[0.5\baselineskip]
W_V(\fragmentKinELab) &= \frac{ w_1^\NucleonSpecies \, \FermiEDiff^2 }
{ \FermiEDiff^2 + (w_2^\NucleonSpecies)^2 }
\\[0.5\baselineskip]
W_D(\fragmentKinELab) &= \frac{ d_1^\NucleonSpecies \, \FermiEDiff^2 }
{ \FermiEDiff^2 + (d_3^\NucleonSpecies)^2 }
\, \exp\!\big( -d_2^\NucleonSpecies \, \FermiEDiff \big)
\\[0.5\baselineskip]
V_{SO}(\fragmentKinELab) &= v_{so1}^\NucleonSpecies
\, \exp\!\big( -v_{so2}^\NucleonSpecies \, \FermiEDiff \big)
\\[0.5\baselineskip]
W_{SO}(\fragmentKinELab) &= \frac{ w_{so1}^\NucleonSpecies \, \FermiEDiff^2 }
{ \FermiEDiff^2 + (w_{so2}^\NucleonSpecies)^2 }
\end{align}
Here the Coulomb contribution to $V_V$ is given by
\begin{equation}
V_\text{Coul} \equiv \delta_{\mathrm{p}\NucleonSpecies}
  \, \overline{V}_{\!C} \, v_1^\mathrm{p} \big( v_2^\mathrm{p}
- 2 \, v_3^\mathrm{p} \, \FermiEDiff + 3 \, v_4^\mathrm{p}
\, \FermiEDiff^2 \big)
\end{equation}
where the symbol $\delta_{\mathrm{p}\NucleonSpecies}$ is defined by
\begin{equation}
\delta_{\mathrm{p}\NucleonSpecies} \equiv
\begin{cases}
0 & \NucleonSpecies = \mathrm{n} \\
1 & \NucleonSpecies = \mathrm{p} \,.
\end{cases}
\end{equation}
Tables 10 and 11 from ref.~\cite{Koning2003} list the parameters needed to
calculate the well depths for a nucleon projectile.

\begin{table}
\vspace{1.0\baselineskip}
\centering
\caption{Radial parameters for the global nuclear optical
potential defined in ref.~\cite{Koning2003}.}
\label{tab:optical_model_params_radial}
\renewcommand{\arraystretch}{1.5}
\begin{tabular}{cc}
\toprule
Parameter & Value (\si{\femto\meter}) \\
\midrule
$r_V$ & $1.3039 - 0.4054\,A^{-1/3}$ \\
$a_V$ & $0.6778 - \num{1.487e-4}\,A$ \\
$r_D$ & $1.3424 - 0.01585\,A^{1/3}$ \\
$a_D^\mathrm{n}$ & $0.5446 - \num{1.656e-4}\,A$ \\
$a_D^\mathrm{p}$ & $0.5187 + \num{5.205e-4}\,A$ \\
$r_{SO}$ & $1.1854 - 0.647\,A^{-1/3}$ \\
$a_{SO}$ & 0.59 \\
$r_C$ & $1.198 + 0.697\,A^{-2/3} + 12.994\,A^{-5/3}$ \\
\bottomrule
\end{tabular}
\end{table}

\subsection{Complex projectiles}

To compute the nuclear optical potential for complex projectiles ($A > 1$),
\marley\ implements a superposition model based on a recommendation by Madland
\cite{Madland1988}. It is equivalent to the default treatment used by TALYS.
Under this approach, the radial optical model parameters for a projectile with
mass number $A$ and proton (neutron) number $Z$ ($N$) are computed by weighting
the corresponding parameters for individual nucleons:
\begin{align}
r_V &= \frac{N \, r_V^n + Z \, r_V^p}{A} & r_D,\text{}r_{SO}
  & \text{ likewise} \\[0.5\baselineskip]
a_V &= \frac{N \, a_V^n + Z \, a_V^p}{A} & a_D,\text{}a_{SO}
  & \text{ likewise.}
\end{align}
The Coulomb radius parameter $r_C$ remains unchanged from the nucleon case.
The well depths are evaluated according to the relations
\begin{align}
V_V(\fragmentKinELab) =& N\,V_V^n(\fragmentKinELab/A)
+ Z\,V_V^p(\fragmentKinELab/A)
\\[0.5\baselineskip] \nonumber
W_V, & W_D \text{ likewise} \\[0.5\baselineskip]
V_{SO}(\fragmentKinELab) =& \frac{V_{SO}^n(\fragmentKinELab)
  + V_{SO}^p(\fragmentKinELab)}{2A}
\\[0.5\baselineskip] \nonumber
W_{SO} &\text{ likewise}
\end{align}
In the expressions above, the superscript $^n$ ($^p$) denotes the value of the
corresponding quantity for an individual neutron (proton), e.g.,
$V_{SO}^n(\fragmentKinELab)$ is the spin-orbit well depth for a neutron
projectile.

\bibliography{marley.bib}

\end{document}